\documentclass{article}

\usepackage{graphicx}
\usepackage{latexsym}
\usepackage{amssymb}

\def\beq{\begin{equation}}
\def\eeq{\end{equation}}
\newcommand{\bea}{\begin{eqnarray}}
\newcommand{\eea}{\end{eqnarray}}

\newcommand{\sgn}{\epsilon}

\def\rightcontract{\mathop{\hbox{\vrule width0.5pt height6pt%
  \vrule height0.5pt width6pt}}}

\begin{document}

\title{\bf \LARGE
Spin-rotation couplings: spinning test particles and Dirac field}

\author{Donato Bini${}^*{}^\S{}^\P$ and
Luca Lusanna$^\P$
\\[4mm]%
{\small \it \thanks{binid@icra.it}\thanks{luca.lusanna@fi.infn.it}~
Istituto per le Applicazioni del Calcolo ``M. Picone'', CNR I-00161 Rome, Italy}\\
{\small \it $^\|$~International Center for Relativistic Astrophysics - I.C.R.A.,}\\
{\small \it University of Rome ``La Sapienza'', I-00185 Rome, Italy}\\%
{\small \it $^\P$~INFN - Sezione di Firenze, Polo Scientifico, via Sansone, 1}\\
{\small \it  I-50019, Sesto Fiorentino (FI), Italy } }

\date{{\small \today }}

\maketitle

\begin{center}
{\it Dedicated to Bahram Mashhoon for his 60th birthday}
\end{center}

\begin{abstract}
The hypothesis of  coupling between spin and rotation introduced
long ago by Mashhoon is examined in the context of \lq\lq 1+3" and \lq\lq 3+1" space-time splitting techniques,
either in special or in general relativity. Its content is discussed in
terms of classical (Mathisson-Papapetrou-Dixon-Souriou model) as well as
quantum physics (Foldy-Wouthuysen transformation for the Dirac field in an
external field), reviewing and discussing all the relevant theoretical 
literature concerning the existence of such
effect. Some original contributions are also included.
\end{abstract}

\vspace*{5mm}

\thispagestyle{empty}

\setcounter{page}{1}

\section{Introduction}

There is a huge literature on the problem of the existence of
spin-rotation couplings either in accelerated frames in Minkowski
space-time or in linearized gravity, due to the absence of
experimental signatures.

Mashhoon was a pioneer in the investigation of the various aspects
of this topic emphasizing the relevance of the {\it locality
hypothesis} in special relativity and of the {\it equivalence
principle} in the comparison of the acceleration and gravity effects
(see e.g. \cite{1,2,3,4,5} for reviews and an exhaustive collection
of references).

The problem of the spin-rotation couplings is mainly formulated in
the context of space-time splitting techniques, namely within the
two different (and competing) points of view termed \lq\lq 1+3" and
\lq\lq 3+1" splittings. 

In the first case ( \lq\lq 1+3" {\it point of
view}), the description of special and general relativistic effects
is done by a time-like observer with an assumed known world line.
As a consequence, the instantaneous 3-space of the observer is
identified with the tangent space of vectors orthogonal to the
observer 4-velocity at each instant of the observer proper time and
usually it is coordinatized with Fermi-like 3-coordinates. In this
way only local information accessible to the observer is used and
this is considered physically acceptable.

However, this description holds only close the observer world line,
because the various instantaneous 3-spaces will intersect each other
at a distance from the world line of the order of the acceleration
lengths \cite{3,5}.
This means that there is neither a {\it globally} defined clock
synchronization convention (replacing Einstein's ${1\over 2}$ one,
valid in special relativistic inertial frames) nor the possibility
to formulate a well-posed Cauchy problem for Maxwell (or Yang-Mills,
or Einstein) equations. Also the type of radar coordinates
introduced in Ref.\cite{6}, based on Einstein convention of clock
synchronization with light signals, have been shown to have similar
limitations\cite{7}. Moreover, within this point of view
to have a consistent definition of instantaneous 3-spaces with an
atlas of coordinates such that the Fermi coordinates are a local
chart around the observer world line is still an open question.

The approach more suited to solve these problems is the \lq\lq
3+1" {\it point of view}, in which one gives a \lq\lq 3+1"  splitting of
Minkowski space-time, namely a nice foliation with space-like
leaves, besides the observer world line; that is one defines a {\it
global non-inertial frame centered on the observer}. Such a
splitting is a {\it generalized clock synchronization convention}:
each space-like leaf is an instantaneous 3-space (in general a
curved Riemannian 3-manifold), which can be used as a Cauchy surface
for field equations. To avoid coordinate singularities like the ones
appearing either with Fermi coordinates or with rotating frames the
foliation has to satisfy the {\it M$\o$ller admissibility
conditions} \cite{8} and its leaves must tend to space-like
hyper-planes at spatial infinity. Then a global non-inertial frame
centered on the given time-like observer can be built by defining
generalized (observer-dependent and Lorentz scalar) radar
4-coordinates $(\tau , \sigma^r)$: the time variable $\tau$,
labeling the simultaneity leaves $\Sigma_{\tau}$, is an arbitrary
monotonically increasing function of the observer proper time;
$\sigma^r$ are curvilinear 3-coordinates on each $\Sigma_{\tau}$
having the world line as origin \footnote{Often we shall use the
notation $\vec \sigma = \{ \sigma^r \}$ for the sake of simplicity.
Moreover, we use a 4-metric with signature $\sgn\, (+---)$, with
$\sgn = \pm$ according to both
standard conventions.}. If $x^{\mu}$ are Cartesian 4-coordinates
in an inertial frame in
Minkowski space-time, 
the coordinate transformation $x^{\mu} \mapsto
\sigma^A = (\tau, \sigma^r)$ has an inverse $\sigma^A \mapsto
x^{\mu} = z^{\mu}(\tau ,\sigma^r)$ defining the embeddings
$z^{\mu}(\tau ,\sigma^r)$ of the simultaneity 3-surfaces. Using the notation
$$z^{\mu}_A = \partial\, z^{\mu}/\partial\, \sigma^A\ ,$$
the induced 4-metric is
$$g_{AB}(\tau,\sigma^r) = \eta_{\mu\nu}\,
z^{\mu}_A(\tau ,\sigma^r)\, z^{\nu}_B(\tau ,\sigma^r)$$ and
M$\o$ller admissibility conditions are given by
 \bea
 && \sgn\, g_{\tau\tau}(\tau ,\sigma^u) > 0,\qquad \sgn\,
 g_{rr}(\tau ,\sigma^u) < 0, \qquad
 \sgn\, {\rm det}\, [g_{rs}(\tau ,\sigma^u)]\, < 0\nonumber \\
 &&{}\nonumber \\
 && {\rm det}
\begin{array}{|ll|}
 g_{rr}(\tau ,\sigma^u)
 & g_{rs}(\tau ,\sigma^u) \\ g_{sr}(\tau ,\sigma^u) &
 g_{ss}(\tau ,\sigma^u) 
\end{array}\, > 0,\qquad \forall r,s \quad {\rm fixed},
\label{1.1}
 \eea
implying ${\rm det}\, [g_{AB}(\tau ,\sigma^u)]\, < 0$.

In Ref.\cite{9} it is shown that with each admissible \lq\lq 3+1"
splitting are associated two congruences of time-like observers (the
natural ones for the given notion of simultaneity):

\noindent i) the Eulerian observers, whose unit 4-velocity field is
the field  of unit normals to the simultaneity surfaces
$\Sigma_{\tau}$;

\noindent ii) the observers whose unit 4-velocity field is
proportional to the evolution vector field of components $\partial
z^{\mu}(\tau ,\sigma^r)/\partial \tau$: in general this congruence
is non-surface forming having a non-vanishing vorticity.

Both the \lq\lq 3+1" and the \lq\lq 1+3"  
point of view may appear not so physical\footnote{Actually, the \lq\lq 3+1" approach requires the knowledge of the data on a whole space-like
hyper-surface which is not factual; similarly the \lq\lq 1+3" is not factual because it requires the knowledge of the data on a whole world line, i.e. also in the \lq\lq future."}, but it allows to arrive at a
well-posed Cauchy problem for field equations, i.e. to a
mathematical control of {\it determinism} once the gauge freedom of
the given field theory has been fixed. Moreover, it allows to
formulate an action principle ({\it parametrized Minkowski
theories}) for every special relativistic isolated system
(particles, strings, fields, fluids) for which a Lagrangian
description is known, such that the transition from an admissible
\lq\lq 3+1" splitting (with associated radar 4-coordinates) to
another one (with new radar 4-coordinates) is obtained using a {\it
gauge transformation}. This property is a consequence of the
invariance of the action under frame-preserving diffeomorphisms and
implies that the physics is independent from the choice of the
synchronization convention, as expected. The same \lq\lq 3+1" point
of view is the starting point of the canonical formulation of metric
and tetrad gravity in globally hyperbolic asymptotically flat (with
suitable boundary conditions ar spatial infinity) space-times, where
there is a notion of global time and where general covariance
(invariance under diffeomorphisms) again implies the gauge
equivalence of the admissible \lq\lq 3+1"  splittings. i.e. of the
global non-inertial frames (the only ones allowed by the {\it
equivalence principle}).

In addition, differently from the Fermi coordinates, it is possible
to give an {\it operational} definition of the generalized radar
4-coordinates. As shown in Refs.\cite{9,10}, given four functions
satisfying certain restrictions due to M$\o$ller conditions, the
on-board computer of a spacecraft may use them to build a grid of
radar 4-coordinates in its future. All these properties are
explained in detail in Refs.\cite{9,10,11} for special relativity
and in Refs.\cite{11,12} for general relativity. Moreover, in
special relativity the restriction of parametrized Minkowski
theories to inertial frames allowed to find the {\it
inertial-rest-frame Wigner-covariant instant form of
dynamics}\cite{11,13}. These results are possible due to a
systematic use of Dirac theory of constraints in the Hamiltonian
description of relativistic systems.

In this paper we mainly reconsider the problem of the spin-rotation
couplings both from the \lq\lq 3+1" and \lq\lq 1+3" points of view,
trying to clarify their interconnections.

\section{Spin-Rotation Couplings in Special Relativity}

\subsection{Parametrized Minkowski Theories and the Rest-Frame
Instant Form of Dynamics.}

{\it Parametrized Minkowski theories}\cite{11,13} have been
developed to describe isolated physical systems in non-inertial
frames in such a way that different conventions for clock
synchronization are connected by gauge transformations.

Given any isolated system admitting a Lagrangian description, one
makes the coupling of the system to an external gravitational field
and then replaces the 4-metric ${}^4g_{\mu\nu}(x)$ with the induced
metric ${}^4g_{AB}[z(\tau ,\sigma^r)]$ associated with an arbitrary
admissible \lq\lq 3+1" splitting. The Lagrangian now depends not
only on the matter configurational variables but also on the
embedding variables $z^{\mu}(\tau ,\sigma^r)$ (whose conjugate
canonical momenta are denoted $\rho_{\mu}(\tau ,\sigma^r)$). Since
the action principle turns out to be invariant under
frame-preserving diffeomorphisms, at the Hamiltonian level there are
four first-class constraints
$${\cal H}_{\mu}(\tau ,\sigma^r) = \rho_{\mu}(\tau ,\sigma^r) -
l_{\mu}(\tau,\sigma^r)\, T^{\tau\tau}(\tau ,\sigma^r) -
z^{\mu}_s(\tau ,\sigma^r)\, T^{\tau s}(\tau ,\sigma^r) \approx 0$$
in strong involution with respect to Poisson brackets 
$$\{ {\cal
H}_{\mu}(\tau ,\sigma^r), {\cal H}_{\nu}(\tau ,\sigma_1^r)\} = 0.
$$

Here $l_{\mu}(\tau ,\sigma^r)$ are the covariant components of the
unit normal to $\Sigma_{\tau}$, while $z^{\mu}_s(\tau ,\sigma^r)$
are the components of three independent vectors tangent to
$\Sigma_{\tau}$. The quantities $T^{\tau\tau}$ and $T^{\tau s}$ are
the components of the energy-momentum tensor of the matter distributed on
$\Sigma_{\tau}$, describing its energy  and momentum  densities. As a
consequence, Dirac's theory of constraints implies that the
configuration variables $z^{\mu}(\tau ,\sigma^r)$ are arbitrary {\it
gauge variables}. Therefore, all the admissible \lq\lq 3+1"
splittings, namely all the admissible conventions for clock
synchronization, and all the admissible non-inertial frames centered
on time-like observers are {\it gauge equivalent}.
By adding four gauge-fixing constraints
$$\chi^{\mu}(\tau ,\sigma^r)
= z^{\mu}(\tau ,\sigma^r) - z^{\mu}_M(\tau ,\sigma^r) \approx 0
$$
($z^{\mu}_M(\tau ,\sigma^r)$ being an admissible embedding),
satisfying the orbit condition
$${\rm det}\, |\{\chi^{\mu}(\tau ,\sigma^r),
{\cal H}_{\nu}(\tau ,\sigma_1^r)\} | \not= 0,$$ we identify the
description of the system in the associated non-inertial frame
centered on some given time-like observer chosen as origin. The
resulting effective Hamiltonian for the $\tau$-evolution turns out
to contain the potentials of the {\it relativistic inertial forces}
present in the given non-inertial frame. Since a non-inertial frame
means the use of its radar coordinates, we see that already in
special relativity {\it non-inertial Hamiltonians are
coordinate-dependent quantities} like the notion of energy density
in general relativity. As a consequence, the gauge variables
$z^{\mu}(\tau ,\sigma^r)$ describe the {\it spatio-temporal
appearances} of the phenomena in non-inertial frames.

Inertial frames centered on inertial observers are a special case of
gauge fixing in parametrized Minkowski theories, where the
embeddings $z^{\mu}(\tau ,\sigma^r)$ are linear in the radar
4-coordinates.

For each configuration of an isolated system there is a special
\lq\lq 3+1" splitting associated with it: the foliation with
space-like hyper-planes orthogonal to the conserved time-like
4-momentum of the isolated system.

This identifies an intrinsic inertial frame, the {\it rest-frame},
centered on a suitable inertial observer (the Fokker-Pryce center of
inertia of the isolated system) and allows to define the {\it
Wigner-covariant rest-frame instant form of dynamics} for every
isolated system \footnote{This happens because in the gauge fixing
use is made of the standard Wigner boost $L^{\mu}_{\nu}(p,
{\buildrel \circ \over p})$ ($p^{\mu}=L^{\mu}_{\nu}(p,{\buildrel
\circ \over p}){\buildrel \circ \over p}^{\nu}$, ${\buildrel \circ
\over p}^{\mu}=\eta \sqrt {\sgn\, p^2} (1;\vec 0 )$, $\eta = sign\,
p^o$).}. Its instantaneous 3-spaces $\Sigma_{\tau}$ are named {\it
Wigner hyper-planes}, because 3-vectors lying into them transform as
Wigner spin-1 3-vectors. See Refs. \cite{14,15} for the development
of a coherent formalism describing all the aspects of relativistic
kinematics for N particle systems, continuous bodies and fields
generalizing all known non-relativistic results:

\noindent i) the classification of the intrinsic notions of
collective variables (canonical non-covariant center of mass;
covariant non-canonical Fokker-Pryce center of inertia;
non-covariant non-canonical M$\o$ller center of energy);

\noindent ii) canonical bases of center-of-mass and relative
variables;

\noindent iii) canonical spin bases and dynamical body-frames for
the rotational kinematics of deformable systems;

\noindent iv) multipolar expansions for isolated systems and their
open subsystems with a Hamiltonian formulation of the
Mathisson-Papapetrou-Dixon-Souriau equations which puts control on their
subsidiary conditions (see the first and third paper in
Ref.\cite{14});

\noindent v) the relativistic theory of orbits \cite{15} (while the
potentials appearing in the energy generator of the Poincare' group
determine the relative motion, the determination of the actual
orbits in the given inertial frame is influenced by the potentials
appearing in the Lorentz boosts);

\noindent vi) the M$\o$ller radius (a classical unit of length
identifying the region of non-covariance of the canonical center of
mass of a spinning system around the covariant Fokker-Pryce center
of inertia; it is an effect induced by the Lorentz signature of the
4-metric; it could be used as a physical ultraviolet cutoff in
quantization).

\noindent vii) the definition of non-inertial rest frames, where the
simultaneity leaves tend to space-like hyper-planes orthogonal to
the total 4-momentum of the system at spatial infinity: only this
family of embeddings is relevant for the \lq\lq 3+1" point of view
of metric and tetrad gravity in globally hyperbolic space-times
\cite{11,12}.

Let us remark that in parametrized Minkowski theories a relativistic
particle with world line $x^{\mu}_i(\tau )$ is described only by the
3-coordinates $\sigma^r = \eta^r_i(\tau )$ defined by
$x^{\mu}_i(\tau ) = z^{\mu}(\tau , \eta^r_i(\tau ))$ and by the
conjugate canonical momenta $\kappa_{ir}(\tau )$. The usual
4-momentum $p_{i\mu}(\tau )$ is a derived quantity satisfying the
mass-shell constraint $\epsilon\, p^2_i = m^2_i$ in the free case.
Therefore, we have a different description for positive and negative
energy particles. All the particles on an admissible surface
$\Sigma_{\tau}$ are simultaneous by construction: {\it this
eliminates the problem of relative times}, which for a long time has
been an obstruction to the theory of relativistic bound states and
to relativistic statistical mechanics (see Ref.\cite{14,15} and its
bibliography for these problems and the related no-interaction
theorem).

\subsection{The Locality Hypothesis and M$\o$ller Conditions}

Let us now consider a class of 4-coordinate transformations
associated with
 the idea of accelerated observers as sequences of
comoving observers  (i.e. the locality hypothesis \cite{9,10}).

The admissible embeddings $x^{\mu} = z^{\mu}(\tau ,\vec \sigma )$,
defined with respect to a given inertial system, must tend to
parallel spacelike hyper-planes at spatial infinity. If $l^{\mu} =
l^{\mu}_{(\infty )}\, {\buildrel {def}\over  =}\,
\epsilon^{\mu}_{\tau}$ [$l^2_{(\infty )} = \sgn$] is the asymptotic
normal, let us define the asymptotic orthonormal tetrad
$\epsilon^{\mu}_A$, $A=\tau ,1,2,3$, by using the standard Wigner
boost for time-like Poincare' orbits $L^{\mu}{}_{\nu}(l_{(\infty )},
{\buildrel \circ \over l}_{(\infty )})$ [${\buildrel \circ \over
l}_{(\infty )}^{\mu} = (1; \vec 0)$]: 
$$\epsilon^{\mu}_A\, {\buildrel
{def}\over =}\, L^{\mu}{}_A(l_{(\infty )}, {\buildrel \circ \over
l}_{(\infty )})$$
with the property $\epsilon^{\mu}_A\,
\eta_{\mu\nu}\, \epsilon^{\nu}_B = \eta_{AB}$ [$= \sgn\, (+---)$].
Then a parametrization of the asymptotic hyper-planes is $z^{\mu} =
x^{\mu}_o + \epsilon^{\mu}_A\, \sigma^A = x^{\mu}(\tau ) +
\epsilon^{\mu}_r\, \sigma^r$ with $x^{\mu}(\tau ) = x^{\mu}_o +
\epsilon^{\mu}_{\tau}\, \tau$ a time-like straight-line (an
asymptotic inertial observer).

Let us define a family of \lq\lq 3+1" splittings of Minkowski
space-time by means of the following embeddings

\bea && z^{\mu}(\tau ,\vec \sigma )= x^{\mu}_o +
\Lambda^{\mu}{}_{\nu}(\tau ,\vec \sigma )\, \epsilon^{\nu}_A\,
\sigma^A\, = {\tilde x}^{\mu}(\tau ) + F^{\mu}(\tau ,\vec \sigma
),\qquad F^{\mu}(\tau ,\vec 0) = 0,\nonumber \\
 &&{}\nonumber \\
 &&{\tilde x}^{\mu}(\tau ) = x^{\mu}_o + \Lambda^{\mu}{}_{\nu}(\tau ,\vec
 0)\, \epsilon^{\nu}_{\tau}\, \tau,  \\
 &&{}\nonumber \\
 && F^{\mu}(\tau ,\vec
 \sigma ) = [\Lambda^{\mu}{}_{\nu}(\tau ,\vec \sigma ) -
 \Lambda^{\mu}{}_{\nu}(\tau ,\vec 0)]\, \epsilon^{\nu}_{\tau}\,
 \tau + \Lambda^{\mu}{}_{\nu}(\tau ,\vec \sigma )\,
 \epsilon^{\nu}_r\, \sigma^r,\nonumber \\
 &&{}\nonumber \\
 &&\Lambda^{\mu}{}_{\nu}(\tau ,\vec \sigma )\, {\rightarrow}_{|\vec
\sigma | \rightarrow \infty}\, \delta^{\mu}_{\nu},\quad
\Rightarrow\,\, z^{\mu}(\tau ,\vec \sigma )\, {\rightarrow}_{|\vec
\sigma | \rightarrow \infty}\,\, x^{\mu}_o + \epsilon^{\mu}_A\,
 \sigma^A = x^{\mu}(\tau ) + \epsilon^{\mu}_r\, \sigma^r,\nonumber
 \label{3.1}
 \eea

\noindent where $\Lambda^{\mu}{}_{\nu}(\tau ,\vec \sigma )$ are
Lorentz transformations ($\Lambda^{\mu}{}_{\alpha}\, \eta_{\mu\nu}\,
\Lambda^{\nu}{}_{\beta} = \eta_{\alpha\beta}$) belonging to the
component connected with the identity of $SO(3,1)$. While the
functions $F^{\mu}(\tau ,\vec \sigma )$ determine the form of the
simultaneity surfaces $\Sigma_{\tau}$, the centroid ${\tilde
x}^{\mu}(\tau )$, corresponding to an {\it arbitrary time-like
observer} chosen as origin of the 3-coordinates on each
$\Sigma_{\tau}$, determines how these surfaces are packed in the
foliation. Since the asymptotic foliation with parallel
hyper-planes, having a constant vector field $l^{\mu} =
\epsilon^{\mu}_{\tau}$ of normals, defines an inertial reference
frame, we see that the foliation (\ref{3.1}) with its associated
non-inertial reference frame is obtained from the asymptotic
inertial frame by means of {\it point-dependent Lorentz
transformations}. As a consequence, the integral lines, i.e. the
non-inertial Eulerian observers, origin of (non-rigid) non-inertial
reference frames, are parametrized as a continuum of comoving
inertial observers as required by the locality hypothesis.

Therefore, in the framework of parametrized Minkowski theories {\it
the locality hypothesis can always be assumed valid modulo gauge
transformations}.

An equivalent parametrization of the embeddings of this family of
reference frames is \bea && z^{\mu}(\tau ,\vec \sigma ) = x^{\mu}_o
+ \epsilon^{\mu}_B\, \Lambda^B{}_A(\tau ,\vec \sigma )\, \sigma^A =
x^{\mu}_o + U^{\mu}_A(\tau ,\vec \sigma )\, \sigma^A
 = {\tilde x}^{\mu}(\tau ) + F^{\mu}(\tau ,\vec \sigma ),\nonumber \\
 &&{}\nonumber \\
 &&{\tilde x}^{\mu}(\tau ) = x^{\mu}_o +
U^{\mu}_{\tau}(\tau ,\vec 0)\, \tau ,\nonumber \\
 &&{}\nonumber \\
 &&F^{\mu}(\tau ,\vec \sigma ) =
 [U^{\mu}_{\tau}(\tau ,\vec \sigma ) -
 U^{\mu}_{\tau}(\tau ,\vec 0)]\, \tau + U^{\mu}_r(\tau ,\vec
 \sigma )\, \sigma^r,
\label{3.2}
 \eea

\noindent where we have defined:

 \bea
 \Lambda^B{}_A(\tau ,\vec \sigma ) &=& \epsilon^B_{\mu}\,
  \Lambda^{\mu}{}_{\nu}(\tau ,\vec \sigma )\,
\epsilon^{\nu}_A,\qquad  U^{\mu}_A(\tau ,\vec \sigma )\,
 \eta_{\mu\nu}\, U^{\nu}_B(\tau ,\vec \sigma ) =
 \epsilon^{\mu}_A\, \eta_{\mu\nu}\,
 \epsilon^{\nu}_B = \eta_{AB}, \nonumber \\
 &&{}\nonumber \\
 U^{\mu}_A(\tau ,\vec \sigma ) &=& \epsilon^{\mu}_B\,
 \Lambda^B{}_A(\tau ,\vec \sigma )\, {\rightarrow}_{|\vec \sigma |
  \rightarrow \infty}\, \epsilon^{\mu}_A,
 \label{3.3}
 \eea

\noindent where $\epsilon^B_{\mu} = \eta_{\mu\nu}\, \eta^{BA}\,
\epsilon^{\nu}_A$ are the inverse tetrads.

A slight generalization of these embeddings allows to find Nelson's
\cite{16} 4-coordinate transformation (but extended from $\vec
\sigma$-independent Lorentz transformations $\Lambda^{\mu}{}_{\nu} =
\Lambda^{\mu}{}_{\nu}(\tau )$ to $\vec \sigma$-dependent ones)
implying M$\o$ller rotating 4-metric\footnote{ $g_{oo} = \sgn [(1+
{{\vec a \cdot \vec x}\over {c^2}})^2 - {{(\omega \times \vec
x)^2}\over {c^2}}]$, $g_{oi} = -\sgn\, {1\over c}\, (\vec \omega
\times \vec x)^i$, $g_{ij} = -\sgn\, \delta_{ij}$, where $\vec a$ is
the time-dependent acceleration of the observer's frame of reference
relative to the comoving inertial frame and $\vec \omega$ is the
time-dependent angular velocity of the observer's spatial rotation
with respect to the comoving frame; $\vec x$ is the position vector
of a spatial point with respect to the origin of the observer's
accelerated frame.}

\bea
 z^{\mu}(\tau ,\vec \sigma ) &=& x^{\mu}_o + \epsilon^{\mu}_A\,
\Big[ \Lambda^A{}_B(\tau ,\vec \sigma )\, \sigma^B + V^A(\tau
,\vec \sigma )\Big],\nonumber \\
 &&{}\nonumber \\
 V^{\tau}(\tau ,\vec \sigma ) &=& \int^{\tau}_o d\tau_1\,
\Lambda^{\tau}{}_{\tau}(\tau_1 ,\vec \sigma ) -
\Lambda^{\tau}{}_{\tau}(\tau ,\vec \sigma )\, \tau,
 V^r(\tau ,\vec \sigma ) \nonumber \\
 &&{}\nonumber \\
&=& \int^{\tau}_o d\tau_1\, \Lambda^r{}_{\tau}(\tau_1, \vec \sigma )
- \Lambda^r{}_{\tau}(\tau ,\vec \sigma )\, \tau .
 \label{3.4}
 \eea

The M$\o$ller conditions (\ref{1.1}) are severe restrictions on the
Lorentz matrices $\Lambda (\tau ,\sigma^r)$, which are stated in
Ref.\cite{9}, where each Lorentz matrix $\Lambda$ is represented as
the product of a Lorentz boost $B$ and a rotation matrix ${\cal R}$
{\it to separate the translational from the rotational effects}
($\vec \beta = \vec v/c$ are the boost parameters, $\gamma (\vec
\beta ) = 1/\sqrt{1- {\vec \beta}^2}$, ${\vec \beta}^2 = (\gamma^2 -
1)/\gamma^2$, $B^{-1}(\vec \beta ) = B(-\vec \beta )$; $\alpha$,
$\beta$, $\gamma$ are three Euler angles and $R^{-1} = R^T$)

\bea
 &&\Lambda (\tau ,\vec \sigma ) = B(\vec \beta (\tau ,\vec \sigma
 ))\, {\cal R}(\alpha (\tau ,\vec \sigma ), \beta (\tau ,\vec
 \sigma ), \gamma (\tau ,\vec \sigma )),\nonumber \\
 &&{}\nonumber \\
 &&B^A{}_B(\vec \beta ) = \left( \begin{array}{cc}
\gamma (\vec \beta )& \gamma (\vec \beta )\, \beta^s\\ \gamma (\vec
\beta )\, \beta^r& \delta^{rs} + {{\gamma^2(\vec \beta )\, \beta^r\,
\beta^s}\over {\gamma (\vec \beta ) + 1}}
\end{array} \right),\nonumber \\
 &&{\cal R}^A{}_B(\alpha ,\beta ,\gamma ) = \left(
 \begin{array}{cc} 1 & 0\\ 0& R^r{}_s(\alpha ,\beta ,\gamma )
 \end{array} \right),
 \label{3.5}
\eea
where $R^r{}_s(\alpha ,\beta ,\gamma )$ is the standard matrix of the Euler angles.

Eqs.(\ref{1.1}) are restrictions on the parameters $\vec \beta (\tau
,\vec \sigma )$, $\alpha (\tau ,\vec \sigma )$, $\beta (\tau ,\vec
\sigma )$, $\gamma (\tau ,\vec \sigma )$ of the Lorentz
transformations, which say that {\it translational accelerations and
rotational frequencies are not independent} but must {\it balance
each other}.

Let us consider two extreme cases.

A)  {\it Rigid non-inertial reference frames with translational
acceleration exist}. An example are the following embeddings, which
are compatible with the locality hypothesis only for $f(\tau ) =
\tau$ (this corresponds to $\Lambda = B(\vec 0)\, {\cal R}(0,0,0)$,
i.e. to an inertial reference frame)

\bea
 &&z^{\mu}(\tau ,\vec \sigma ) =x^{\mu}_o +
\epsilon^{\mu}_{\tau}\, f(\tau ) + \epsilon^{\mu}_r\,
\sigma^r,\nonumber \\
 &&{}\nonumber \\
 &&g_{\tau\tau}(\tau ,\vec \sigma ) = \sgn\,
 \Big({{d f(\tau )}\over {d\tau}}\Big)^2,\quad g_{\tau r}(\tau ,\vec \sigma )
 =0,\quad g_{rs}(\tau ,\vec \sigma ) = -\sgn\, \delta_{rs}.
 \label{3.6}
 \eea

This is a foliation with parallel hyper-planes centered on a
centroid $x^{\mu}(\tau ) = x^{\mu}_o + \epsilon^{\mu}_{\tau}\,
f(\tau )$ (origin of 3-coordinates). The hyper-planes have
translational acceleration ${\ddot x}^{\mu}(\tau ) =
\epsilon^{\mu}_{\tau}\, \ddot f(\tau )$, so that they are not
uniformly distributed like in the inertial case $f(\tau ) = \tau$.

B) On the other hand  {\it rigid rotating reference frames do not
exist}. Let us consider the embedding (compatible with the locality
hypothesis) with $\Lambda = B(\vec 0)\, {\cal R}(\alpha (\tau
),\beta (\tau ),\gamma (\tau ))$ and $x^{\mu}(\tau ) = x^{\mu}_o +
\epsilon^{\mu}_{\tau}\, \tau$, which corresponds to a foliation with
parallel to space-like hyper-planes with normal $l^{\mu} =
\epsilon^{\mu}_{\tau}$. It can be verified that it is not the
inverse of an admissible 4-coordinate transformation, because the
associated $g_{\tau\tau}(\tau ,\vec \sigma )$ has a zero at $ \sigma
= \sigma_R $ such that \footnote{We use the notations $\vec \sigma =
\sigma\, \hat \sigma$, $\sigma = |\vec \sigma |$, $\vec \Omega =
\Omega\, \hat \Omega$, ${\hat \sigma}^2 = {\hat \Omega}^2 = 1$,
$\Omega^u = - {1\over 2}\, \epsilon^{urs}\, (\dot R\,
R^{-1})^r{}_s$, $b^{\mu}_r(\tau ) = \epsilon^{\mu}_s\, R_r{}^s(\tau
)$.}

\beq \sigma_R = {1\over {\Omega (\tau )}}\, \Big[- {\dot x}
_{\mu}(\tau )\, b^{\mu}_r(\tau )\, (\hat \sigma \times \hat
\Omega(\tau ))^r + \sqrt{{\dot x}^2(\tau ) + [{\dot x}_{\mu}(\tau
)\, b^{\mu}_r(\tau )\, (\hat \sigma \times \hat \Omega (\tau
))^2}]^2 \,\, \Big],
 \label{3.7}
 \eeq

\noindent with $\quad \sigma_R \rightarrow \infty$ for $\Omega
\rightarrow 0$. At $\sigma = \sigma_R$ the time-like vector
$z^{\mu}_{\tau}(\tau ,\vec \sigma )$ becomes light-like (the
so-called {\it horizon problem} of the rotating coordinate systems),
while for an admissible foliation with space-like leaves it must
always remain time-like.

As shown in Ref.\cite{9}, the simplest notion of simultaneity
compatible with the locality hypothesis when rotations are present
is obtained with embeddings where there is a rotation matrix $R(\tau
,|\vec \sigma|)$, namely the rotation varies as a function of some
radial distance $|\vec \sigma |$ ({\it differential rotation}) from
the arbitrary time-like world line $x^{\mu}(\tau )$, origin of the
3-coordinates on the simultaneity surfaces. Since the 3-coordinates
$\sigma^r$ are Lorentz scalar we shall use the radial distance
$\sigma = |\vec \sigma| = \sqrt{\delta_{rs}\, \sigma^r\, \sigma^s
}$, so that $\sigma^r = \sigma \, {\hat \sigma}^r$ with
$\delta_{rs}\, {\hat \sigma}^r\, {\hat \sigma}^s = 1$. These
embeddings are

\bea
 && z^{\mu}(\tau ,\vec \sigma ) = x^{\mu}(\tau ) + \epsilon^{\mu}_r\,
R^r{}_s(\tau , \sigma )\, \sigma^s\,
 {\buildrel {def}\over =}\, x^{\mu}(\tau ) + b^{\mu}_r(\tau
 ,\sigma )\, \sigma^r,\nonumber \\
 &&{}\nonumber \\
 &&R^r{}_s(\tau ,\sigma ) {\rightarrow}_{\sigma \rightarrow
 \infty} \delta^r_s,\qquad \partial_A\, R^r{}_s(\tau
 ,\sigma )\, {\rightarrow}_{\sigma \rightarrow
 \infty}\, 0,\nonumber \\
 &&{}\nonumber \\
 &&b^{\mu}_s(\tau ,\sigma ) = \epsilon^{\mu}_r\, R^r{}_s(\tau
 ,\sigma )\, {\rightarrow}_{\sigma \rightarrow
 \infty}\, \epsilon^{\mu}_s,\quad [b^{\mu}_r\, \eta_{\mu\nu}\, b^{\nu}_s](\tau ,\sigma )
 = - \sgn\, \delta_{rs},\nonumber \\
 &&{}\nonumber \\
 && R(\tau ,\sigma ) = R(\alpha(\tau ,\sigma ), \beta (\tau ,\sigma
 ), \gamma (\tau ,\sigma )),\nonumber \\
 &&{}\nonumber \\
 && \alpha (\tau ,\sigma) =F(\sigma )\, \tilde \alpha (\tau
 ),\quad \beta (\tau ,\sigma ) = F(\sigma )\, \tilde \beta (\tau
 ),\quad \gamma (\tau ,\sigma )=F(\sigma )\, \tilde \gamma (\tau ).
 \label{3.8}
 \eea

Since $z^{\mu}_r(\tau ,\vec \sigma ) = \epsilon^{\mu}_s\,
\partial_r\, [R^s{}_u(\tau ,\sigma )\, \sigma^u]$, it follows that
the normal to the simultaneity surfaces is $l^{\mu} =
\epsilon^{\mu}_{\tau}$, namely the hyper-surfaces are {\it parallel
space-like hyper-planes}. These hyper-planes have translational
acceleration ${\ddot x}^{\mu}(\tau )$ (it could be simulated with a
rigid boost) and a rotating 3-coordinate system with rotational
frequency $\Omega^r(\tau ,\sigma ) = - {1\over 2}\, \epsilon^{ruv}\,
\Big[ R^{-1}(\tau ,\sigma )\, {{\partial R(\tau ,\sigma )}\over
{\partial \tau}}\Big]^{uv}\, {\rightarrow}_{\sigma \rightarrow
\infty}\, 0$.

As shown in Ref.\cite{9}, the M$\o$ller conditions (\ref{1.1}) imply

 \beq
 0< F(\sigma ) < \frac{1}{M\,\sigma},\qquad
 {{d F(\sigma )}\over {d \sigma }} \not= 0,\quad
 \mbox{ or }\qquad| \Omega^r(\tau ,\sigma ) | < {{m}\over { K\,
   \sigma}}\,(K-1),
 \label{3.9}
 \eeq

\noindent where the constants $m > 0$ and $K > 1$ are determined by
the 4-velocity ${\dot x}^{\mu}(\tau )$ of the observer [$m ={\rm  min}\,\{
\sgn\, {\dot x}_{\mu}(\tau )\, l^{\mu}\}$, $|\vec v(\tau )| \leq
\sgn\, {\dot x}_{\mu}(\tau )\, l^{\mu}/ K$ with $\vec v(\tau)$ the
observer 3-velocity].
 \medskip

Every function $F(\sigma )$ satisfying Eq.(\ref{3.9}) gives rise to
a M$\o$ller-admissible  non-inertial rotating frame. As said in
Subsections C and D of Section VI of Ref.\cite{10}, the choice
$$F(\sigma ) = \displaystyle {{1 + {{\omega^2\, R^2}\over {c^2}}}\over {1 +
{{\omega^2\, \sigma^2}\over {c^2}}}} < \displaystyle  {2\over {1 + {{\omega^2\,
\sigma^2}\over {c^2}}}} < {1\over {\omega\, \sigma}},\quad F(\sigma
)\, \rightarrow_{c\rightarrow \infty}\, 1 + {{\omega^2\, (R^2 -
\sigma^2)}\over {c^2}} + O({1\over {c^4}}),$$
 replaces the rigid
rotation $\Omega (\sigma ) = \omega$ for $\sigma < R$ of a rotating
disk of radius $R$ (with $\omega\, R < c$) with an admissible
differential rotation $\Omega (\sigma ) = \omega\, F(\sigma )$. By
varying the admissible functions $F(\sigma )$ (a gauge
transformation in  parametrized Minkowski theories) we can
approximate the step function $\Omega (\sigma ) = \omega$ for
$\sigma < R$, $\Omega (\sigma ) = 0$ for $\sigma > R$, as much as we
wish.

In Ref.\cite{10} there is also the treatment of the Sagnac effect in
this framework.

This means that, while the linear velocities ${\dot x}^{\mu}(\tau )$
and the translational accelerations ${\ddot x}^{\mu}(\tau )$ are
arbitrary, the allowed rotations $R(\alpha ,\beta ,\gamma )$ on the
leaves of the foliation {\it have the rotational frequencies},
namely the angular velocities $\Omega^r(\tau ,\sigma )$, {\it
limited by an upper bound proportional to the minimum of the linear
velocity $v_l(\tau ) = \sgn\, {\dot x}_{\mu}(\tau )\, l^{\mu}$
orthogonal to the parallel hyper-planes}.

In Refs.\cite{9,10} it is shown that if we consider the observers of
the second skew congruence associated with these embeddings, whose
unit 4-velocity is $V^{\mu}_{\tau}(\tau ,\sigma^r) =
z^{\mu}_{\tau}(\tau ,\sigma^r)/\sqrt{\sgn\, g_{\tau\tau}(\tau
,\sigma^r)}$, and we endow them with an ortho-normal tetrad
$V^{\mu}_A(\tau ,\sigma^s) = \Big(V^{\mu}_{\tau}(\tau ,\sigma^s);
V^{\mu}_r(\tau ,\sigma^s)\Big)$, then we can study $d\,
V^{\mu}_A(\tau ,\sigma^s)/d\tau$. For each value of $\sigma^r$,
namely for each observer of the congruence, we obtain an
acceleration matrix and the associated {\it acceleration lengths}.
But now, differently from the case of Fermi coordinates, the radar
4-coordinates associated with these M$\o$ller-admissible embeddings
are globally defined and do not develop coordinate singularities at
a distance from the observer world line of the order of the
accelerations lengths.

\subsection{Quantum Mechanics in Non-Inertial Frames}

The postulates of non-relativistic quantum mechanics are formulated
in {\it global inertial reference frames}, connected by the
transformations of the  kinematical (extended) Galilei group, which,
due to the {\it Galilei relativity principle}, relate the
observations of an inertial observer to those of another one. The
self-adjoint operators on the Hilbert space, in particular the {\it
Hamiltonian operator} (governing the time-evolution in the
Schroe\-dinger equation and identified with the {\it energy
operator} in the projective representation of the quantum Galilei
group associated with the system), correspond to the quantization of
classical quantities defined in these frames. The resulting quantum
theory is extremely successful {\it both for isolated and open
systems} (viewed as sub-systems of isolated systems).

At the relativistic level conceptually nothing changes: we have the
{\it relativity principle} stating the impossibility to distinguish
special relativistic inertial frames and the kinematical Poincare'
group replacing the Galilei one. Again the {\it energy} is one of
the generators of the kinematical group and is identified with the
{\it canonical Hamiltonian} governing the evolution of a
relativistic Schroedinger equation.

In this framework, with a semi-relativistic treatment of the
electro-magnetic field we get an extremely successful theory of
atomic spectra in inertial reference frames both for isolated
inertial atoms (closed systems) and for accelerated ones in presence
of external forces (open systems). The following cases are an
elementary list of possibilities.\medskip

a) {\it Isolated atom} - From the time-dependent Schroedinger
equation $i\, {{\partial}\over {\partial t}}\, \psi = H_o\, \psi$,
through the position $\psi = e^{i\, E_n\, t/\hbar }\, \psi_n$ we get
the time-independent Schroe\-din\-ger equation $H_o\, \psi_n = E_n\,
\psi_n$ for the stationary levels and the energy spectrum $E_n$ with
its degenerations. Being isolated the atom can decay only through
spontaneous emission.

b) {\it Atom in an external c-number, maybe time-dependent,
electro-magnetic field} - Now the (energy) Hamiltonian operator is
in general non conserved (open system). Only for time-independent
external fields it is clear how to define the time-independent
Schroedinger equation for the stationary states and the
corresponding (modified) spectrum. Time-independent external
electro-magnetic fields lead to removal of degeneracies (Zeeman
effect) and/or shift of the levels (Stark effect). With
time-dependent external fields we get the Schroedinger equation $i\,
{{\partial}\over {\partial t}}\, \psi = H(t)\, \psi$ with $H(t) =
H_o + V(t)$. Therefore at each instant $t$ the self-adjoint operator
$H(t)$ defines a different basis of the Hilbert space with its
spectrum, but, since in general we have $[H(t_1), H(t_2)] \not= 0$,
it is not possible to define a unique associated  eigenvalue
equation and an associated spectrum varying continuously in $t$.
Only when we have $[H(t_1), H(t_2)] = 0$ we can write $H(t)\,
\psi_n(t) = E_n(t)\, \psi_n(t)$ with time-dependent eigenvalues
$E_n(t)$ and a visualization of the spectrum as a continuous
function of time. In any case, when $V(t)$ can be considered a
perturbation, time-dependent perturbation theory with suitable
approximations can be used to find the transition amplitudes among
the levels of the unperturbed Hamiltonian $H_o$. Now the atom can
decay both for spontaneous or stimulated emission and be excited
through absorption.

c) {\it Atom plus an external c-number \lq\lq mechanical" potential
inducing, for instance, the rotational motion of an atom fixed to a
rotating platform} (see the Moessbauer effect) - If the c-number
potential is $V(t)$, i.e. it is only time-dependent, we have $i\,
{{\partial}\over {\partial t}}\, \psi = [H_o + V(t)]\, \psi = H(t)\,
\psi$ with $[H(t_1), H(t_2)] = 0$ and the position $\psi = e^{i\,
\int_o^t\, V(t_1)\, dt_1/\hbar}\, \psi_1$ leads to $i\,
{{\partial}\over {\partial t}}\, \psi_1 = H_o\, \psi_1$, so that the
energy levels are $E_{1n} = E_n + \int_o^t dt_1\, V(t_1)$. The
addition of a c-number external time-dependent electro-magnetic
field leads again to the problems of case b).

d) At the relativistic level we can consider the {\it isolated
system atom + electro-magnetic field} as an approximation to the
theory of bound states in quantum electrodynamics. Both the atom and
the electro-magnetic field are separately accelerated open
subsystems described in an inertial frame.

In any case the modifications of the energy spectrum of the isolated
atom is induced by {\it physical force fields} present in the
inertial frame of the observer.

In case c) we can consider an accelerated observer carrying a
measuring apparatus and rotating with the atom with the theory of
measurement based on the {\it locality hypothesis}. As a consequence
the observer will detect the same spectrum as an inertial observer.

Let us consider the description of the previous cases from the point
of view of a {\it non-inertial observer} carrying a measuring
apparatus by doing a {\it passive coordinate transformation} adapted
to the motion of the observer. Since, already at the
non-relativistic level, there is  {\it no relativity principle for
non-inertial frames}, there is {\it no kinematical group} (larger
than the Galilei group) whose transformations connect the
non-inertial measurements to the inertial ones:  given the
non-inertial frame with its linear and rotational accelerations with
respect to a standard inertial frame, we can only define the {\it
succession} of time-dependent Galilei transformations identifying at
each instant the {\it comoving inertial observers}, with the same
measurements of the non-inertial observer if the locality hypothesis
holds.

Since we are considering a {\it purely passive viewpoint}, there is
no physical reason to expect that the atom spectra will change:
there are no physical either external or internal forces but only a
different viewpoint which changes the appearances and introduces the
fictitious (or inertial) mass-proportional forces to describe these
changes.

In the framework of parametrized Minkowski theories, like in general
relativity, the passive frame-preserving diffeomorphisms on
Minkowski space-time imply a special relativistic form of general
covariance, but {\it do not form} a kinematical group (extending the
Poincare' group), because there is {\it no relativity principle} for
non-inertial observers. Therefore, there is {\it no kinematical
generator interpretable as a non-inertial energy}.

The $c \rightarrow \infty$ limit of parametrized Minkowski theories
allows to define {\it parametrized Galilei theories} and to describe
non-relativistic congruences of non-inertial observers. Again there
is no relativity principle for such observers, no kinematical group
extending the Galilei one and, therefore, no kinematical generator
to be identified as a non-inertial energy.

All the existing attempts \cite{17} to extend the standard
formulation of quantum mechanics from global rigid inertial frames
to  {\it special global rigid non-inertial reference frames} carried
by observers with either linear (usually {\it constant})
acceleration or rotational (usually {\it constant}) angular velocity
are equivalent to the definition of suitable {\it time-dependent
unitary transformations} acting in the Hilbert space associated with
inertial frames.

While in inertial frames the generator of the time evolution, namely
the Hamiltonian operator $H$ appearing in the Schroedinger equation
$i\, {{\partial}\over {\partial t}}\, \psi = H\, \psi$, also
describes the energy of the system, after a time-dependent unitary
transformation $U(t)$ the {\it generator $\tilde H(t) = U(t)\, H\,
U^{-1}(t) + i \dot U(t)\, U^{-1}(t)$ of the time evolution}
\footnote{This {\it non-inertial Hamiltonian} containing the
potential $i \dot U\, U^{-1}$ of the {\it fictitious or inertial
forces is not a generator of any kinematical group}.} in the
transformed Schroedinger equation $i\, {{\partial}\over {\partial
t}}\, \tilde \psi = \tilde H(t)\, \tilde \psi$, with $\tilde \psi =
U(t)\, \psi$, {\it differs from the energy operator} $H^{'} = U(t)\,
H\, U^{-1}(t)$. And also in this case like in example b), only if we
have $[\tilde H(t_1), \tilde H(t_2)] = 0$ it is possible to define a
unique stationary equation with time-dependent eigenvalues for
$\tilde H(t)$.

The situation is analogous to the Foldy-Wouthuysen transformation
\cite{18}, which is a time-dependent unitary transformation when it
exists: in this framework $H^{'}$ is the energy, while $\tilde H(t)$
is the Hamiltonian for the new Schroedinger equation and the
associated S-matrix theory (theoretical treatment of
semi-relativistic high-energy experiments, $\pi\, N$,..).

Since in general the self-adjoint operator $\tilde H(t)$ does not
admit a unique associated eigenvalue equation \footnote{Even when it
does admit such an equation, we have $< \psi | H | \psi
> \not= < \tilde \psi | \tilde H | \tilde\psi
>$ and different stationary states are connected, following the
treatment of the time-independent examples given by Kuchar (quoted
in Refs.\cite{17}) in which it is possible to find the spectrum of
both of them , by a generalized transform.} and, moreover, since the
two self-adjoint operators $\tilde H $ and $H^{'}$ are in general
(except in the static cases) non commuting, there is {\it no
consensus} about the results of measurements in non-inertial frames:
{\it does a non-inertial observer see a variation of the emission
spectra of atoms}? Which is the spectrum of the hydrogen atom seen
by a non-inertial observer? Since for constant rotation we get
$\tilde H = H^{'} + \vec \Omega \cdot \vec J$, does the uniformly
rotating observer see the inertial spectra or are they modified by a
Zeeman effect? If an accelerated observer would actually measure the
Zeeman levels with an energy measurement, this would mean that the
stationary states of $\tilde H$ (and not those of the inertial
energy operator $H^{'}$) are the relevant ones. Proposals for an
experimental check of this possibility are presented in
Ref.\cite{19}. Usually one says that a possible non-inertial Zeeman
effect from constant rotation is either too small to be detected or
masked by physical magnetic fields, so that the distinction between
$\tilde H$ and $H^{'}$ is irrelevant from the experimental point of
view.

Here we have exactly the same problem like in the case of an atom
interacting with a time-dependent external field: the atom is
defined by its inertial spectrum, the only one  unambiguously
defined when $[\tilde H(t_1), \tilde H(t_2)] \not= 0$. When
possible, time-dependent perturbation theory is used to find the
transition amplitudes among the inertial levels. Again only in
special cases (for instance time-independent $\tilde H$) a spectrum
for $\tilde H$ may be evaluated and usually, except in special cases
like the Zeeman effect, it has no relation with the inertial
spectrum (see Kuchar in Ref.\cite{17} for an example). Moreover,
also in these special cases the two operators may not commute so
that the two properties described by these operators cannot in
general be measured simultaneously.

As a consequence of these problems the description of measurements
in non-inertial frames is often replaced by an explanation of how to
correlate the phenomena to the results of measurements of the energy
spectra in inertial frames. For instance in the Moessbauer effect
one  only considers the correction for Doppler effect (evaluated by
the instantaneous comoving inertial observer) of unmodified spectra.
Regarding the spectra of stars in astrophysics, only correction for
gravitational red-shift of unmodified spectra are considered. After
these corrections the inertial effects connected with the emission
in non-inertial frames manifest themselves only in a {\it
broadening} of the inertial spectral lines. In conclusion {\it atoms
are always identified through their inertial spectra in absence of
external fields}. The non-inertial effects, precluding the unique
existence of a spectrum continuous in time, are usually small and
appear as a noise over-imposed to the continuous spectrum of the
center of mass.

An apparatus for measuring $\tilde H$ can be an {\it interferometer}
measuring the variation $\triangle\, \phi$ of the phase of the
wave-function describing the two wave-packets propagating, in
accordance with the non-inertial Schroedinger equation (one uses the
Dirac-Feynman path integral with $\tilde H$ to evaluate $\triangle\,
\phi$) along the two arms of the interferometer. However, {\it the
results of the interferometer only reveal the eventual non-inertial
nature of the reference frame}, namely  {\it they amount to a
detection of the non-inertiality of the frame of reference}, as
remarked in Ref.\cite{9}. In this connection see Ref.\cite{20} on
neutron interferometry, where there is a full account of the status
of the experiments for the detection of inertial effects.

Now in the non-relativistic literature there is an {\it active
re-interpretation in terms of gravitational potentials of the
previous passive view} according to a certain reading of the {\it
non-relativistic limit of the classical (weak or strong) equivalence
principle} (universality of free fall or identity of inertial and
gravitational masses) and to its extrapolation to quantum mechanics
(see for instance Hughes \cite{21} for its use done by Einstein).
According to this interpretation, at the classical level the {\it
passive fictitious forces} seen by the accelerated observer are
interpreted as an {\it active external Newtonian gravitational
force}  acting in an inertial frame, so that at the quantum level
$\tilde H$ is interpreted as the energy operator in an inertial
frame in presence of an external quantum gravitational potential
$\tilde H - H^{'} = i\, \dot U U^{-1}$. Therefore the shift from the
levels of $H^{'}$ to those of $\tilde H$ is justified and expected.
However this interpretation and use of the equivalence principle is
subject to criticism already at the classical level.

A first objection is that a physical external gravitational field
(without any connection with non-inertial observers) leads to the
Schroedinger equation $i\, {{\partial}\over {\partial t}}\, \psi =
[H + V_{grav}]\, \psi$ and not to $i\, {{\partial}\over {\partial
t}}\, \tilde \psi = [UHU^{-1} + V_{grav}]\, \tilde \psi$, $\tilde
\psi = U\, \psi$.

A further objection is that the interpretations based upon the use
of the equivalence principle in the special relativistic treatment
of atomic, nuclear and particle physics  rely on Einstein's
statements, which are explicitly  referred to {\it static constant
gravitational fields}. But according to Synge \cite{22}, static
fields does not exists in general relativity: only tidal effects
exist.

In Ref.\cite{23} there is the quantization of relativistic scalar
and spinning particles described by means of parametrized Minkowski
theories in a class of non-inertial frames (like the ones of
Eq.(\ref{3.8})), whose embeddings have the parametrization

\bea
 z^{\mu}(\tau ,\vec \sigma ) &\approx& \theta (\tau )\, {\hat
 U}^{\mu}(\tau ) + \epsilon^{\mu}_a(\hat U(\tau ))\, {\cal
 A}^a(\tau ,\vec \sigma ) \nonumber \\
&&{}\nonumber \\
&=&
 x^{\mu}_U(\tau ) + \epsilon^{\mu}_a(\hat U(\tau ))\, \Big[{\cal
A}^a(\tau ,\vec
\sigma ) - {\cal A}^a(\tau ,\vec 0)\Big] ,\nonumber \\
 &&{}\nonumber \\
x^{\mu}_U(\tau ) &=&  z^{\mu}(\tau ,\vec 0) = \theta (\tau )\, {\hat
U}^{\mu}(\tau )
 + \epsilon^{\mu}_a(\hat U(\tau ))\, {\cal A}^a(\tau ,\vec
 0).
  \label{3.10}
  \eea

\noindent The simultaneity surfaces $\Sigma_{\tau}$ are space-like
hyper-planes orthogonal to the arbitrary time-like unit vector
${\hat U}^{\mu}(\tau )$. We see that  $\theta (\tau )$ {\it
describes the freedom in the choice of the mathematical time $\tau$,
$\partial^2 {\cal A}^a(\tau ,\vec 0)/\partial \tau^2$ the arbitrary linear
3-acceleration of the non-inertial frame and ${{\partial {\cal
A}^a(\tau ,\vec \sigma )}\over {\partial \tau}} - {{\partial {\cal
A}^a(\tau ,\vec 0 )}\over {\partial \tau}}$ its angular velocity, describing the arbitrary
admissible differentially rotating 3-coordinates. As a consequence,
a gauge fixing for $\theta (\tau )$ and ${\cal A}^a(\tau ,\vec
\sigma )$ realizes the choice of a well defined non-inertial frame}
centered on the time-like observer $x^{\mu}_U(\tau )$.

The positive-energy particles on $\Sigma_{\tau}$ are described by
canonically conjugate 3-vectors ${\vec \eta}_i(\tau )$, ${\vec
\kappa}_i(\tau )$ with the particle world lines given by
$x^{\mu}_i(\tau ) = z^{\mu}(\tau ,{\vec \eta}_i(\tau ))$ ($\sgn\,
p_i^2 = m^2_i$). The effective non-inertial Hamiltonian $H_{ni}$ in
one of the previous non-rigid non-inertial frames is given in
Eq.(2.47) of Ref.\cite{23}. By means of a time-dependent canonical
transformation (point in the momenta) one gets (see Eq.(4.7) of
Ref.\cite{23}) the form\footnote{$\kappa^{'}_{i\, a}(\tau ) = {\cal
A}^s_a(\tau ,{\vec \eta}_i(\tau ))\, \kappa_{i s}(\tau )$, where
${\cal A}^s_a(\tau ,\sigma^u)$ is the matrix inverse of
${{\partial\, {\cal A}_a(\tau ,\sigma^u)}\over {\partial\,
\sigma^r}}$.}
$$H_{ni} = H_{inertial} + \sum_{i=1}^N\,
\Big[ \vec v(\tau ) + \Omega(\tau, {\vec \eta}_i(\tau )) \times
{\vec \eta}_i(\tau )\Big] \cdot {\vec \kappa}^{'}_i(\tau ).$$

The quantization is based on a multi-temporal quantization scheme
for systems with first-class constraints, in which only the particle
degrees of freedom $\eta^r_i(\tau )$, $\kappa_{ir}(\tau )$ are
quantized. The gauge variables, describing the appearances (inertial
effects) of the motion in non-inertial frames, are treated as
c-numbers (like the time in the Schroedinger equation with a
time-dependent Hamiltonian) and the physical scalar product does not
depend on them. The resulting Schroedinger equation in a given
non-rigid non-inertial frame of the given class is $i\, \hbar\,
{{\partial}\over {\partial\, \tau}}\, \psi (\tau , {\vec \eta}_i) =
{\hat H}_{ni}\, \psi (\tau ,{\vec \eta}_i)$ (see Eq.(3.71) of
Ref.\cite{23}), with the self-adjoint non-inertial Hamiltonian
corresponding to a particular ordering in the quantization of the
classical $H_{ni}$. Since the previously quoted time-dependent
canonical transformation is point in the momenta it becomes a
time-dependent unitary transformation at the quantum level,
connecting the inertial Hamiltonian (the energy) to the non-inertial
one as expected.

With this type of relativistic kinematics it has been possible to
separate the center of mass \footnote{At the relativistic level this
is done with a canonical transformation which is {\it point only} in
the momenta \cite{14,15}.} and to verify that the spectra of
relativistic bound states in non-inertial frames are not modified by
inertial effects.

The non-relativistic limit \cite{24} allows to recover the few
existing attempts of quantization in non-inertial frames as
particular cases. Now it is possible to restrict the theory to rigid
non-inertial frames, where one gets for the non-inertial Hamiltonian
$$H_{ni} = \sum_i\, {{{\vec p}_i^2(t)}\over {2m_i}} - \vec v(t) \cdot
{\vec P}(t) - \vec \omega (t) \cdot \vec J(t)$$ (see Eq.(4.8) of
Ref.\cite{24}).

Therefore, the standard total angular momentum -rotation coupling
(but not an angular momentum -linear acceleration one) emerges as
the potential of inertial forces from the quantization in rigid
non-inertial frames. For spinning particles one gets the
spin-rotation coupling, because the spin is included in the total
angular momentum. However, as already said, the eigenvalues of the
non-inertial Hamiltonian are a measure of non-inertiality and not of
energy. At the relativistic level only non-rigid non-inertial frames
are M$\o$ller admissible and this implies that the spin-rotation
coupling is replaced by a much more complex potential for the
relativistic inertial forces. The same complications happen also in
non-relativistic non-rigid non-inertial frames.

The main open problem is the quantization of the scalar Klein-Gordon
field (and of every other field) in non-inertial frames, due to the
Torre and Varadarajan \cite{25} no-go theorem, according to which in
general the evolution from an initial space-like hyper-surface to a
final one is {\it  not unitary} in the Tomonaga-Schwinger
formulation of quantum field theory. From the \lq\lq 3+1" point of
view there is evolution only among the leaves of an admissible
foliation and the possible way out from the theorem lies in the
determination of all the admissible \lq\lq 3+1" splittings of
Minkowski space-time satisfying the following requirements: i)
existence of an instantaneous Fock space on each simultaneity
surface $\Sigma_{\tau}$ (i.e. the $\Sigma_{\tau}$'s must admit a
generalized Fourier transform); ii) unitary equivalence of the Fock
spaces on $\Sigma_{\tau_1}$ and $\Sigma_{\tau_2}$ belonging to the
same foliation (the associated Bogoljubov transformation must be
Hilbert-Schmidt), so that the non-inertial Hamiltonian is a
Hermitean operator; iii) unitary gauge equivalence of the \lq\lq
3+1" splittings with the Hilbert-Schmidt property. The overcoming of
the no-go theorem would help also in quantum field theory in curved
space-times and in condensed matter (here the non-unitarity implies
non-Hermitean Hamiltonians and negative energies).

\subsection{Spinning Particles in non-Inertial Frames: the
Foldy-Wouthuysen Transformation and  Quantization}

The manifestly Lorentz covariant description of spinning particles
in inertial frames at the pseudo-classical level is based on an
action principle in which there are Grassmann variables
$\xi_i^{\mu}$, $\xi_{i5}$, i=1,..,N, for the description of their
spin structure \cite{26}. While we have $\xi^{\mu}_i\, \xi^{\nu}_i +
\xi_i^{\nu}\, \xi_i^{\mu} = 0$, the Grassmann variables describing
the spin of different particles are assumed to commute:
$\xi^{\mu}_i\, \xi^{\nu}_j = \xi^{\nu}_j\, \xi^{\mu}_i$. After
quantization the Grassmann variables of each particle generate the
Clifford algebra of Dirac matrices: $\xi_{i5} \mapsto
\sqrt{{{\hbar}\over 2}}\, \gamma_5$, $\xi^{\mu}_i \mapsto
\sqrt{{{\hbar}\over 2}}\, \gamma_5\, \gamma^{\mu}$. The spin tensor
of each particle is $S^{\mu\nu}_i = - i\, \xi^{\mu}_i\,
\xi^{\nu}_i$. Besides suitable second class constraints, there are
two first class constraints associated with each particle:
$\chi_{Di} = p_{i\mu}\, \xi^{\mu}_i - m_i\, \xi_{i5} \approx 0$,
$\chi_i = p^2_i - m^2_i \approx 0$ with the property $\{ \chi_{Di},
\chi_{Dj} \} = i\, \delta_{ij}\, \chi_i$. After quantization we have
$\chi_{Di} \approx 0 \,\,\mapsto \gamma_5\, (p_{i\mu}\, \gamma^{\mu}
- m_i)\, \psi(p_i) = 0$, while the mass shell constraints become the
Klein-Gordon equation implied by the square of the Dirac equation.
See the second review in Refs.\cite{11} for all the applications of
pseudo-classical mechanics.

In particular in Ref.\cite{27} there is the definition of the
pseudo-classical Foldy-Wouthuysen (FW) transformation as a canonical
transformation generated by a function $S_{cl} = 2i\, {\vec p} \cdot
{\vec \xi}\, \xi_5\, \theta (\vec p)$ ($tg\, 2\, |\vec p|\, \theta
(\vec p) = |\vec p|/m$), which after quantization becomes the
unitary FW transformation $e^{i\, S}$, $S = \beta\, \vec \alpha
\cdot \vec p\, \theta (\vec p)$ sending the Dirac Hamiltonian $H =
\vec \alpha \cdot \vec p + \beta\, m$ into $e^{i\, S}\, H\, e^{-i\,
S} = \beta\, \sqrt{m^2 + {\vec p}^2}$. In Ref.\cite{27} it was
possible with his technique to find an exact FW transformation for
an electron interacting with a non-homogeneous magnetic field.

In Ref.\cite{13} there was the reformulation of N charged scalar
particles interacting with the electromagnetic field in the
framework of parametrized Minkowski theories. Now only
positive-energy scalar particles are described and the
regularization of Coulomb self-energies was possible in the
semiclassical approximation of using Grassmann-valued electric
charges ($Q^2_i = 0$, $Q_i\, Q_j = Q_j\, Q_i \not= 0$ for $i \not=
j$). After quantization each Grassmann-valued charge gives rise to a
two-level system (charge 0 and charge $e$ or charge $- e$ and charge
$e$). A Shanmugadhasan canonical transformation adapted to the first
class constraints produced the description of the system in the
radiation gauge for the electro-magnetic field, containing only
transverse vector potential and electric field. In Ref.\cite{28} the
Lienard-Wiechert solution in the rest-frame instant form was found.
Then in Ref.\cite{29} it was possible to identify the semi-classical
Darwin potential among the N charged scalar particles, a result that
till now had been obtained only coming down from QFT through the
Bethe-Salpeter equation and its instantaneous approximations.

Then in Ref.\cite{30} there was the reformulation of N spinning
particles in the framework of parametrized Minkowski theories, as
the pseudo-classical basis of the positive energy $({1\over 2}, 0)$
part of the $({1\over 2}, {1\over 2})$ solutions of the Dirac
equation. Essentially one has to eliminate the variables $\xi_{i5}$
from  the model of Ref.\cite{26} and add by hand the new constraints
$\phi_i \approx p_{s\mu}\, \xi^{\mu}_i \approx 0$, where the
4-momentum $p_{s\mu}$ is weakly equal to the total conserved
4-momentum of the N particle system. The global conditions $\phi_i
\approx 0$ \footnote{All the simultaneity surface $\Sigma_{\tau}$ is
involved since $p_{s\mu} = \int d^3\sigma\, \rho_{\mu}(\tau
,\sigma^r)$ with $\rho_{\mu}(\tau ,\sigma^r)$ being the momentum
conjugate to the embedding $z^{\mu}(\tau ,\sigma^r)$.} turn out to
belong to the set of second class constraints so that the spin
structure of the positive-energy spinning particles can be described
only by means of a Wigner spin-1 Grassmann 3-vectors ${\vec {\tilde
\xi}}_i$ (the spin vector of each particle is ${\vec S}_i = -
{i\over 2}\, {\vec {\tilde \xi}}_i \times {\vec {\tilde \xi}}_i$),
which after quantization become the Pauli matrices: ${\vec {\tilde
\xi}}_i \mapsto \sqrt{{{\hbar}\over 2}}\, \vec \sigma$. The
resulting theory is the relativistic version of the non-relativistic
Pauli equation. This is the formalism used in Ref.\cite{23} for the
quantum mechanics of spinning particles in non-inertial frames. Then
in Ref.\cite{30} there is the coupling of N charged (with
Grassmann-valued charges) spinning particles to the electro-magnetic
field and the determination of the Lienard-Wiechert solution on the
Wigner hyper-planes of the rest-frame instant form.

Finally in Ref.\cite{31} there was the determination of the
Darwin-Salpeter potential for N positive-energy spinning particles
in the scheme developed in Ref.\cite{30}. However in Ref.\cite{30}
it was not clear whether extra coupling of the spin to the electric
field were present for positive energy spinning particles.
Therefore, the pseudo-classical FW transformation of Ref.\cite{27}
(in this paper the electric charge was not Grassmann-valued) was
applied to positive-energy spinning particles with Grassmann-valued
charges interacting with an arbitrary external electro-magnetic
field. For a single particle the final constraint (whose
quantization would produce the relativistic Pauli equation) is \bea
p_o &\approx& Q\, A_o(x) + {{Q\, \vec p \cdot \vec E(x) \times \vec
S}\over {(m + \sqrt{m^2 + {\vec p}^2})\, \sqrt{m^2 + {\vec p}^2}}} +\nonumber \\
&& +\sqrt{m^2 + \Big(\vec p - Q\, \vec A(x)\Big)^2 + 2\, Q\, \vec S
\cdot \vec B(x)}. \eea This result, valid in inertial frames, was
incorporated in the formalism of Ref.\cite{30}: after re-expressing
the theory as a parametrized Minkowski theory, there was the
restriction to the electro-magnetic radiation gauge and then to the
rest-frame instant form on Wigner hyper-planes. The methods of
Ref.\cite{29} allowed to find the Darwin-Salpeter inter-particle
potential and the correct spin-orbit and spin-spin interactions.

Most of these results have been obtained in the inertial rest-frame
instant form, but, as discussed in the previous Subsection, their
extension to non-inertial frames \cite{23} can be done with
time-dependent unitary transformations at the quantum level (for now
in absence of the electro-magnetic field due to the
Torre-Varadarajan no-go theorem for field theories on space-like
hyper-surfaces). The non-rigidity of relativistic non-inertial
frames replaces the spin-rotation couplings with more complicated
inertial effects.

Notice that in Ref.\cite{10}, following the preliminary results of
Ref.\cite{13}, there is a study of Maxwell theory as a parametrized
Minkowski theory. There are indications that in uniformly rotating
(M$\o$ller forbidden) non-inertial frames the non-inertial electric
and magnetic fields (used for the magnetosphere of pulsars
\cite{32,33}) are not connected to the inertial ones by Lorentz
transformations. A better understanding and the connection with
Mashhoon's nonlocal electrodynamics \cite{34} is needed.

\subsection{The Multipolar Expansion and the Mathis\-son-Papa\-petrou-Dixon-Sou\-riau Equations}

In Refs.\cite{29,31} there was the evaluation of the energy-momentum
tensor of the isolated system of N positive-energy either scalar or
spinning particles plus the electro-magnetic field in the radiation
gauge on the Wigner hyper-planes of the inertial rest-frame instant
form. Then the description of Dixon multipolar expansion in the
rest-frame instant form done in the third paper of Ref.\cite{14}
allows to study the Mathisson-Papapetrou-Dixon-Souriau pole-dipole equations
(see the next Section) in a consistent Hamiltonian framework both
for isolated systems and for their open subsystems. For open
subsystems the subsidiary conditions needed by these equations are
automatically selected by the Hamiltonian formalism, once a choice
is made for the collective variable to be used as a centroid in the
pole-dipole approximation of the extended open subsystem. The
drawback of these multipolar expansions is that any set of variables
one chooses to describe the centroid (monopole), the spin (dipole)
and the higher multi-poles do not form a canonical basis of the
phase space: their complicated Poisson brackets reduce the utility
of the multipolar approximation. Moreover the Hamiltonian
description of extended systems is strictly speaking equivalent to
the multipolar expansion only when the energy-momentum is an
analytic function of its variables \cite{dixonjmp67}.

\subsection{The Dirac Field in Non-Inertial Frames}

In Ref.\cite{35} there is the reformulation of Dirac fields as
parametrized Minkowski theories with a special emphasis on the study
of the highly non-trivial algebra of first and second class
constraints \footnote{Second class constraints appear because the
Dirac equation is a first order partial differential equation.} on
arbitrary admissible simultaneity space-like 3-surfaces. A Grassmann
valued Dirac spinor was used, so that after quantization one can
recover the anti-commuting Dirac fields of QFT. However the same
algebra of constraints is obtained for the Dirac equation in first
quantization. The starting point was a special form of the
Lagrangian, which is possible only in the flat Minkowski space-time
where the 4-spin connection vanishes (see Appendix A of
Ref.\cite{35}). After the analysis of the constraints in
non-inertial frames, the main results are stated in the inertial
rest-frame instant form. However the determination of the
non-inertial Hamiltonian in a fixed admissible \lq\lq 3+1" splitting
should be obtainable with a time-dependent unitary transformation
following the scheme of Ref.\cite{23}. This would be the \lq\lq 3+1"
point of view on the Dirac equation in non-inertial frames to be
compared with the results (valid locally around the accelerated
observer) of the 1+3 point of view quoted in Refs.\cite{3} (see in
particular the results of Hehl and Ni \cite{36}). Again the
spin-rotation couplings should become complicated inertial effects
depending on the chosen non-rigid non-inertial frame.

\section{Spin-Rotation Couplings in General Relativity}

\subsection{The 1+3 Splitting of the Space-Time}

In general relativity a reference frame is defined by a congruence
of time-like curves, say the set of the world lines of a family of
observers. We denote by $u$ the unit tangent vector ($u\cdot u =-1$)
of the world lines of a generic reference frame. The splitting of
the space-time along $u$ and its orthogonal local rest space $LRS_u$
gives the measurement relative to $u$ of any tensor field defined on
a domain of the space-time; similarly one can obtain the formulation
relative to $u$ of any tensor equation \footnote{In this and the next sections units are chosen here in order
that the speed of light in empty space satisfies $c=1$; 
moreover the metric signature conventions is fixed assuming  $\sgn=-1$ whereas, for notations and
conventions,  we follow \cite{manyfaces}.}.

The measurement of the space-time metric gives rise to the spatial
metric $P(u)_{\alpha\beta}=g_{\alpha\beta}+u_\alpha u_\beta $, i.e.,
the spatial projection operator with respect to $u$; the temporal
projection operator along $u$ is $-u\otimes u$. Analogously
$\eta(u)_{\alpha\beta\sigma}=u^\rho \eta_{\rho\alpha\beta\sigma}$ is
the only spatial field resulting from the measurement of the unit
(oriented) volume 4-form $\eta$; it defines the spatial cross
product $\times_u$ as well as the spatial dual operation on $LRS_u$.

From the measurement of a $p$-form $S$ only two distinct fields
result: the purely spatial or ``magnetic" part $S(u)^{(M)}$ (a
$p$-form) and the ``electric" part $S(u)^{(E)}$ (a $(p-1)$-form),
the spatial projection of any contraction with a single factor of
$u$:
\begin{equation}
S=u^\flat\wedge S(u)^{(E)} +S(u)^{(M)}\ ,
\end{equation}
where the completely covariant (contravariant) form of a generic
tensor $X$ is denoted by $X^\flat$ ($X^\sharp$).

For a generic 2-form $S$ the component notation for this
decomposition is
 \begin{equation}
S_{\alpha\beta} = 2 u_{[\alpha}S(u)^{(E)}{}_{\beta ]} +
S(u)^{(M)}{}_{\alpha\beta}= 2 u_{[\alpha}S(u)^{(E)}{}_{\beta ]} +
 \eta(u)_{\alpha\beta}{}^\sigma S(u)^{(M)}{}_\sigma ,
 \end{equation}
where $S(u)^{(M)}{}_\sigma$ denotes the spatial dual of
$S(u)^{(M)}{}_{\alpha\beta}$.

According to the measurement process, the spatial and temporal
projection of space-time derivative operators gives rise to the
corresponding spatial and temporal counterparts. As described in
\cite{manyfaces} the spatial covariant derivative is defined as
$\nabla_{(u)}{}_\alpha=P(u)P(u)_\alpha^\beta \nabla_\beta$ (the
first projection operator acts on the tensorial indices of the field
after the derivative is applied) and the spatial Lie derivative
along a generic field $X$ is $\pounds_{(u)}{}_X=P(u)\pounds_X$.
Similarly one can construct  temporal derivatives: the Lie temporal
derivative $\nabla_{({\rm lie},u)}=P(u)\pounds_u$, the Fermi-Walker
temporal derivative $\nabla_{({\rm fw},u)}=P(u)\nabla_u$.

The covariant derivative of $u$ has the following decomposition:
 \begin{equation}
 \nabla_\alpha u^\beta  =
 \eta(u)_\alpha {}^\beta{}_\mu \omega(u)^\mu
 +\theta(u)_{\alpha}{}^{\beta}
 -u_\alpha a(u){}^\beta \ ,
 \end{equation}
where $a(u)= \nabla_{({\rm fw},u)} u \in LRS_u$ is the acceleration
vector, $\theta(u)$ $\in LRS_u\otimes LRS_u$ is the (symmetric)
deformation 2-tensor and $\omega(u)\in LRS_u$ is the vorticity
vector of the observer congruence.

If $X$ is a tensor field defined only along the line $\ell_U$,
(parametrized by the proper time $\tau_U$ and having unit tangent
$U$), then the ``measurement" of $\frac{DX}{d\tau_U}$, intrinsic
derivative along $U$, by a family of observers with 4-velocity $u$
is:
\begin{equation}
\frac{D_{({\rm fw},U,u)} }{d\tau_{(U,u)}}X=[\nabla_{({\rm
fw},u)}+\nabla_{(u)}{}_{\nu_{(U,u)}}]X \ ,
\end{equation}
where $U=\gamma(U,u)[u+\nu(U,u)]$,
$d\tau_{(U,u)}=\gamma(U,u)d\tau_U$ is the differential of the
standard relative time parametrization along the line $U$, and the
field $X$ on the right hand side is some smooth extension of $X$ to
a neighborhood of the line.

It is convenient to introduce the composition of projection maps
from the local rest space of an observer onto that of another:
$$P(U,u)=P(U)P(u): LRS_u \longrightarrow LRS_U\ .$$
We also introduce the relative gravitational field $F^{(G)}_{({\rm
fw},U,u)}$ (see \cite{manyfaces}):
\begin{eqnarray}
F_{({\rm fw},U,u)}^{(G)}
&=&-\gamma(U,u)^{-1}P(u)\frac{Du}{d\tau_U}=-P(u)\frac{Du}{ d\tau_{(U,u)} }
\nonumber \\
&=& \gamma(U,u)\left[g(u)+ ||\nu(U,u)|| \left( \frac12
\hat\nu(U,u)\times_u H(u) +\right.\right. \nonumber \\
&&\left. \left. -\theta(u)\rightcontract \hat\nu (U,u)
\right)\right]\ ,
 \label{eq:FGFW}
\end{eqnarray}
where $g(u)=-a(u)$ represents the gravito-electric field while
$H(u)=2\omega(u)$ is the gravito-magnetic field; $\rightcontract $
denotes right-contraction.

The last equation of (\ref{eq:FGFW}) gives the gravitational force a
Lorentz-like form allowing the introduction of terms such as
gravito-electromagnetic force and gravito-electromagnetism.

The Riemann tensor is represented by its ``electric" (${\cal
E}(u){}_{\alpha\beta}$), ``magnetic" (${\cal H}(u){}_{\alpha\beta}$)
and  ``mixed" (${\cal F}(u){}_{\alpha\beta}$) parts, namely
  \begin{eqnarray}
 {\cal E}(u){}_{\beta\delta} & = & R_{\alpha\beta\gamma\delta}
 u^\alpha u^\gamma \ ,\nonumber \\
 {\cal H}(u){}_{\beta\delta} & = & -\frac{1}{2}\eta(u)^{\gamma\mu}{}_\delta
 R_{\alpha\beta\gamma\mu}u^\alpha \ , \nonumber  \\
 {\cal F}(u){}_{\beta\delta} & = & \frac{1}{4}\eta(u)^{\alpha\mu}{}_\beta
 \eta(u)^{\gamma\sigma}{}_\delta R_{\alpha\mu\gamma\sigma}.
 \end{eqnarray}

This summarizes the essential \lq\lq 1+3" splitting formalism. 
It will be used in the next section.

\subsection{Spinning Particles in External Gravitational Fields}

\subsubsection{ Extended Bodies in Classical General Relativity}

In general relativity, an extended body is described by its
associated energy momentum tensor. A small body can be studied by a
multi-pole expansion method: the body is equivalently described by a
set of multi-pole (energy) moments defined along a central line
\cite{math37,papa51,dixon64,dixon69,dixon70}. The cutoff at successive
multi-pole orders defines a hierarchy of elementary multi-pole
particles (see e.g.~\cite{math37,papa51,dixon74,fer94}). The first step is
the point particle (or single-pole), governed by the geodesic
equation of motion. The second one is the dipole (``spinning")
particle which interests us here. The equations of motion for such a
particle were first derived in the pure gravitational case by
Papapetrou \cite{papa51}:

\begin{eqnarray}
\displaystyle\frac{D}{d\tau_U}p^\alpha & = &
-\frac12 R^\alpha{}_{\beta\rho\sigma}U^\beta  S^{\rho\sigma}  \nonumber\\
&&\nonumber\\
 \displaystyle\frac{D}{d\tau_U} S^{\alpha\beta}& = & p^\alpha U^\beta-p^\beta
 U^\alpha ;
 \label{papa}
\end{eqnarray}
where $R_{\alpha\beta\rho\sigma}$ is the Riemann tensor, $p^\alpha$
is the (generalized) momentum vector, $S^{\alpha\beta}$ is a
(antisymmetric) spin tensor, $U=DX/d\tau_U$ is the unit tangent
vector ($U^\alpha U_\alpha =-1$) of the ``center line" $\ell_U$ used
to make the multi-pole reduction, and where $X=X(\tau_U)$ is the
center point whose world line is $\ell_U$. The fields $S$, $U$ and
$p$ are defined only along $\ell_U$. 

The case in which both gravitational and electromagnetic fields are
present was studied by Dixon and Souriau \cite{dixon74,souriau74}.

It is well known that the number of independent equations in
(\ref{papa}) is less than that of the unknown quantities: 3
additional scalar supplementary conditions (SC) are needed for the
scheme to be completed. Once a suitable choice has been made,
$\ell_U$, $p$ and $S$ can, in principle, be determined by the
equations. The various supplementary conditions which are considered
in the literature are all of the form ${\hat u}^\alpha
S_{\alpha\beta}=0$ for some  time-like unit vector $\hat u$ along
the world line $\ell_U$. According to the special relativistic
analogy (\cite{MTW}, p. 161), this is equivalent to defining the
central line $\ell_U$ as the world line of the centroid of the body
with respect to an observer family with 4-velocity $\hat{u}$.

The supplementary conditions most frequently discussed in the
literature are
\begin{description}
\item{(CP)}  $\hat{u}=u$ (Corinaldesi-Papapetrou condition: see e.g.
\cite{cori51,barker74}), where $u$ is a (known) preferred family of
observers usually suggested by the background;
\item{(T)}  $\hat{u}=p/||p||={\bar u}$
(Tulkzyjew's condition: see e.g.
\cite{tulc59,dixon64,dixon74,ehlers77});
\item{(P)} $\hat{u}=U$ (Pirani's
condition: see e.g. \cite{pir56,ragusa95}).
\end{description}
Clearly the fields $\ell_U$, $p$ and $S$ all depend on the choice of
supplementary conditions \cite{ehlers77} so a more precise notation
would be: $X_{(SC)}$, $U_{(SC)}$, $p_{(SC)}$, $S_{(SC)}$, where the
index values $SC=CP,T,P$ correspond to these choices. This
cumbersome notation (but not confusion) is usually avoided.

It is clear from the above discussion that the most widely 
accepted description of spinning test particles in relativity 
is far from being complete, at least for the arbitrary choice of 
supplementary conditions required to make the model compatible.

Fortunately, the Hamiltonian methods of the previous Section, when suitably
extended to include gravity, will allow the general relativistic
canonical formulation of these results.
In the first of Ref. \cite{14} there is a detailed special relativistic study 
of the possible effective center of motion of an open subsystem of an isolated system.
It turns out that at the hamiltonian level the most convenient  choices are either the M$\o$ller energy center of motion or
the Tulkzyjew's one.

\subsubsection{The Mathisson-Papapetrou-Dixon-Souriau Equations of Motion}

The Papapetrou-Dixon-Souriau equations of motion of a spinning test
particle in a given gravitational and electromagnetic background
\cite{souriau70,souriau70bis,souriau72,souriau74} are given by:

\begin{eqnarray}
\displaystyle\frac{D}{d\tau_U}p^\alpha & = & -\frac12
R^\alpha{}_{\beta\rho\sigma} S^{\rho\sigma}U^\beta
+eF^\alpha{}_\beta U^\beta  -\frac12 \lambda S^{\mu\nu}
\nabla^\alpha F_{\mu\nu}  \ ,\cr &&\cr
 \displaystyle\frac{D}{d\tau_U} S^{\alpha\beta}& = & p^\alpha U^\beta-p^\beta
 U^\alpha +\lambda [S^{\alpha\mu}F_\mu{}^\beta-
S^{\beta\mu}F_\mu{}^\alpha] \ ,
  \label{sou}
\end{eqnarray}
where $F^{\alpha\beta}$ is the electromagnetic field and  $\lambda$
is an electromagnetic coupling scalar.

The spatial dual of the spin-electromagnetism coupling term
$S^{\alpha\mu}F_\mu{}^\beta-S^{\beta\mu}F_\mu{}^\alpha $ appearing
in the second of equations (\ref{sou}) coincides with the coupling
term found by Bargman, Michel and Telegdi \cite{BMT}. The classic
Papapetrou's scheme is obtained from (\ref{sou}) by assuming $F=0$,
i.e. by neglecting the electromagnetic field.

The various terms arising from the coupling with the 
spin of the gravitational, electromagnetic and inertial fields 
can be derived after a systematic use of splitting techniques.

Precisely, it is convenient to introduce the following notation for the
spin-gravity and spin-electromagnetism coupling terms:
 \begin{eqnarray}
 {\cal R}_{\alpha\beta} &=&\frac12  R_{\alpha\beta\mu\nu} S^{\mu\nu} \ ,\nonumber \\
Q_\alpha & = & \frac12 S^{\mu\nu}\nabla_\alpha F_{\mu\nu} \ ,\nonumber \\
N^{\alpha\beta} &=&
S^{\alpha\mu}F_\mu{}^\beta-S^{\beta\mu}F_\mu{}^\alpha\ .
 \end{eqnarray}

Assume a) $p=||p||\bar u=M_0 \bar u $, defined along $\ell_U$, is
time-like: ${\bar u}^\alpha {\bar u}_\alpha =-1$ and b) $U$ and
$\bar u$ may be extended in a regular way in a neighborhood of the
line $\ell_U$. We then have two time-like congruences, $\bar u$ and
$U$, at our disposal.

Both $\bar u$ and $U$ are associated with the particle in a natural
way and so are both candidates for defining the proper rest frame of
the particle. There is no agreement in the literature about which of
them should define the proper rest frame. This also leads to the
problem of definition of the center of mass world line as the
P-center or the T-center respectively.

Let us now consider a generic observer field $u$. The splitting of
$U$ and $p$ along $u$ and $LRS_u$ is:
\begin{eqnarray}
 U&=&\gamma (U,u)[u+\nu(U,u)], \cr
 p&=&||p||\bar u=E(p,u)u +p(u)=M_0 \gamma(\bar u, u) [u +\nu(\bar u,u)],
\label{splitUP}
\end{eqnarray}
where $E(p,u)=\gamma(\bar u ,u)||p||=\gamma(\bar u ,u)M_0$. Let us
introduce the following notation for the electric and magnetic parts
of the antisymmetric 2-tensor fields $S$, $F$, $N$ and ${\cal R}$
which appear in the equations: $L(u) = S(u)^{(E)}$, $S(u) =
S(u)^{(M)}$, $E(u)=F(u)^{(E)}$, $B(u)=F(u)^{(M)}$,
$N(u)=N(u)^{(E)}$, $M(u)=N(u)^{(M)}$, $K(u) = {\cal R}(u)^{(E)}$,
$T(u)= {\cal R}(u)^{(M)}$, and extending the notation to the 1-form
$Q$, $Q(u)=Q(u)^{(M)}$, $q(u)=Q(u)^{(E)}$: $Q=Q(u)+q(u)u, \qquad
q(u)=-u\cdot Q\ $. In particular, we have:
\begin{eqnarray}
 K(u)^\alpha &=& {\cal E}(u)^\alpha{}_{\sigma}L(u)^\sigma +
 {\cal H}(u)^\alpha{}_{\sigma}S(u)^\sigma \ ,\cr
 T(u)^\alpha &=& {\cal F}(u)_\mu{}^\alpha S(u)^\mu
-{\cal H}(u)_\mu{}^\alpha L(u)^\mu \ ,\cr N(u)^\alpha &=&
[E(u)\times_u S(u)+B(u)\times_u L(u)]^\alpha\ ,\cr M(u)^\alpha &=&
[B(u)\times_u S(u)+L(u)\times_u E(u)]^\alpha\ ,\cr Q(u)^\alpha &=&
S(u)^\sigma \nabla_{(u)}{}^\alpha B(u)_\sigma -
L(u)^\sigma \nabla_{(u)}{}^\alpha E(u)_\sigma\nonumber \\
&&+ [N(u)\times_u\omega(u)]^\alpha +\theta(u)^\alpha{}_\mu N(u)^\mu
\ ,\cr q(u)&=& L(u)^\sigma \nabla_{({\rm fw},u)}E(u)_\sigma -
S(u)^\sigma \nabla_{({\rm fw},u)} B(u)_\sigma -N(u)\cdot a(u)\ .
\label{KH}
\end{eqnarray}
The background is assumed to be completely known; in particular the
electromagnetic field satisfies the Maxwell equations. It is also
useful to write the following decomposition for $K(U)\in LRS_U$:
 $$
 K(U)=P(u,U)K(U)+u[\nu(U,u)\cdot P(U,u)K(U)]\ .
 $$
We have $P(u,U)K(U)=\gamma(U,u)[K(u)+ \nu(U,u)\times_u T(u)]\ .$

With these definitions, the equations of motion (\ref{sou}) are
equivalent to the following set:

\begin{eqnarray}
\gamma(U,u) \frac{D_{({\rm fw},U,u)}}{d\tau _{(U,u)}} [E(p,u)\nu
(\bar u,u)] &=& E(p,u)F^{(G)}_{({\rm fw},U,u)}
+P(u,U)[-K(U)+eE(U)]\nonumber \\
&&-\lambda Q(u) \ ,\label{splitp}\\
\gamma(U,u) \frac{D_{({\rm fw},U,u)}}{d\tau _{(U,u)}} E(p,u) &=&
E(p,u)\nu(\bar u,u)\cdot
F^{(G)}_{({\rm fw},U,u)} \nonumber \\
&&+\nu(U,u)\cdot P(U,u)[-K(U)+eE(U)]-\lambda q(u) \ ,\label{splitE}\\
\gamma(U,u) \frac{D_{({\rm fw},U,u)}}{d\tau _{(U,u)}} L(u)&=&
S(u)\times_u F^{(G)}_{({\rm fw},U,u)}
+\gamma(U,u)E(p,u)[\nu(U,u)-\nu(\bar u,u)]\nonumber \\
&& +\lambda N(u) \ ,\label{splitL}
\\
\gamma(U,u) \frac{D_{({\rm fw},U,u)}}{d\tau _{(U,u)}} S(u)
&=&F^{(G)}_{({\rm fw},U,u)}\times_u L(u)
+\gamma(U,u)E(p,u)\nu(\bar u,u)\times_u\nu(U,u)\nonumber \\
&& +\lambda M(u)\ . \label{splitS}
\end{eqnarray}

The reference frame field $u$ appearing in eqs
(\ref{splitp})--(\ref{splitS}) is a generic one. If one specializes
it to $U$, following simplifications occur:
 \begin{eqnarray}
 & &D_{({\rm fw},U,U)}/d\tau _{(U,U)}=\nabla_{({\rm fw},U)}\ , \nonumber\\
 & &\nu(U,U)=0\ ,\qquad \gamma(U,U) =1 \ ,
 \nonumber\\
 & &F^{(G)}_{({\rm fw},U,U)}=-a(U) \ ,\nonumber
 \end{eqnarray}
so that we have

\begin{eqnarray}
 \nabla_{({\rm fw}, U)}[E(p,U)\nu(\bar u,U)] &=&
 -E(p,U)a(U)-K(U)+eE(U)-\lambda Q(U) \ ,\nonumber \\
 \nabla_{({\rm fw}, U)} E(p,U) &=& -E(p,U)\nu(\bar u,U)\cdot a(U)
-\lambda q(U) \ ,\nonumber\\
 \nabla_{({\rm fw}, U)} L(U)&=& -S(U)\times_U a(U)
-E(p,U)\nu (\bar u,U)+\lambda N(U) \ ,\nonumber\\
 \nabla_{({\rm fw}, U)} S(U) &=&-a(U)\times_U
L(U) +\lambda M(U) \ . \label{splitSU}
\end{eqnarray}
If we instead specialize the reference frame field to $\bar u$, one
has $\nu(\bar u,\bar u)=0$ so that

\begin{eqnarray}
M_0 F^{(G)}_{({\rm fw},U,\bar u)}&=&
-P(\bar u,U)[-K(U)+eE(U)]+\lambda Q(\bar u) \ ,\label{splitpb}\\
 \gamma(U,\bar u)D_{({\rm fw},U,\bar u)}/d\tau _{(U,\bar u)} M_0 &=&
\nu(U,\bar u)\cdot P(\bar u,U)[-K(U)+eE(U)]-\lambda q(\bar u) \
,\nonumber \\
&&{}
\label{splitEb}\\
\gamma(U,\bar u) D_{({\rm fw},U,\bar u)}/d\tau _{(U,\bar u)} L(\bar
u) &=& S(\bar u)\times_{\bar u} F^{(G)}_{({\rm fw},U,\bar u)}
+\gamma(U,\bar u)E(p,\bar u)\nu(U,\bar u)\nonumber \\
&&+\lambda N(\bar u) \ ,\label{splitLb} \\
 \gamma(U,\bar u)D_{({\rm fw},U,\bar u)}/d\tau _{(U,\bar u)} S(\bar u) &=&
F^{(G)}_{({\rm fw},U,\bar u)}\times_{\bar u} L(\bar u) +\lambda
M(\bar u)\ .
 \label{splitSb}
 \end{eqnarray}
Taking into account equation (\ref{splitpb}), one can rewrite
equation (\ref{splitEb}) as an energy theorem:
\begin{equation}
 \gamma(U,\bar u)D_{({\rm fw},U,\bar u)}/d\tau _{(U,\bar u)} M_0 =
-M_0 \nu(U,\bar u)\cdot F^{(G)}_{({\rm fw},U,\bar u)} -\lambda
[Q(\bar u)\cdot\nu(U,\bar u)+q(\bar u) ]. \label{splitEb2}
\end{equation}
The motion of the body relative to $u$ can in principle be described
by either the spatial velocity $\nu(U,u)$ or the generalized one
$\nu(\bar u,u)$, the two descriptions being inequivalent. From
ordinary relativistic kinematics we have the following ``addition of
velocity law"
\begin{equation}
\frac{\gamma(U,\bar u)}{\gamma(U, u)}\nu(U,\bar u)= P(\bar u,
u)[\nu(U,u)-\nu(\bar u,u)]\ .
\end{equation}

Analogously, the acceleration of the particle relative to $u$ can
also be described by the Fermi-Walker spatial derivative of either
$\nu(U,u)$ or $\nu(\bar u,u)$, and again the two are inequivalent.
The generalized acceleration corresponding to $\nu(\bar u,u)$ is
given by equation (\ref{splitp}). On the right hand side of
(\ref{splitp}) are terms corresponding to the gravitational field,
the electromagnetic field, and a sum of spin-gravity-
electromagnetism coupling terms. It is convenient to introduce the
following (relative) tidal acceleration, due entirely to the spin of
the particle, to represent this coupling:

\begin{eqnarray}
\gamma(U,u) A(\bar u, u) &=& -P(u,U)K(U)-\lambda Q(u)\nonumber \\
& = &  \gamma(U,u) D_{({\rm fw},U,u)}/d\tau _{(U,u)} [E(p,u)\nu(\bar
u,u)] \nonumber \\&& -E(p,u)F^{(G)}_{({\rm fw},U,u)}-eP(u,U)E(U) \ .
\label{oldcoupling}
\end{eqnarray}

The choice of the supplementary condition is fundamental: 1) it
defines the world line $\ell_U$, support for all the fields we are
dealing with; 2) the equations are not formally invariant for
different choices. In fact, from (\ref{splitUP}) in the P,T,CP cases
we have respectively
\begin{eqnarray}
\hbox{\rm (P):} & \qquad L(u)=S(u)\times_u \nu(U,u)\ , \label{psc}\\
\hbox{\rm (T):} & \qquad L(u)=S(u)\times_u \nu(\bar u,u)\label{tsc} \ ,\\
\hbox{\rm (CP):} & \qquad L(u)=0 \label{cpsc}\ .
\end{eqnarray}
When substituted into (\ref{splitp}) these three conditions lead to
a total of 6 different expressions for the sum of the spin-gravity
and the spin-electromagnetism couplings.

\subsubsection{ Pseudo-Classical Mechanics}

In the \lq\lq 3+1" framework of pseudo-classical mechanics \cite{26}, instead, a study
was made of the coupling of a spinning particle to an external
gravitational field in Ref.\cite{37}. It was shown that he algebra
of constraints closes consistently only is the external
gravitational field is {\it torsionless}. The Pauli-Lubanski spin
4-vector $\Sigma^{\mu}$ of the spinning particle satisfies the
equation of motion ${{d\, \Sigma^{\mu}(\tau )}\over {d\tau}} +
\Gamma^{\mu}_{\lambda\nu}\, {\dot x}^{\nu}\, \Sigma^{\lambda} = 2\,
g^{\mu\rho}\, R_{\rho\nu}\, {\dot x}^{\nu}\, \Gamma_5$ ($\Gamma_5 =
\xi_o\, \xi_1\, \xi_2\, \xi_3 \mapsto \gamma_5$)): the spin vector
is not parallel transported where the Ricci tensor is non-zero. This
an analogue of the Bargmann-Michel-Telegdi equation in external
electro-magnetic fields \cite{BMT}. After quantization the Dirac
equation has the standard form $(i\, \hbar\, \gamma^{\mu}\,
\nabla_{\mu} - m)\, \psi (x) = 0$.

\subsection{The Rest-Frame Instant Form of metric and tetrad
Gravity in the 3+1 Point of View.}

In Ref.\cite{12}, after a study of the constraints of ADM canonical
metric gravity, the same was done for ADM tetrad gravity, the theory
needed for the coupling of fermions to gravity.

The Hamiltonian formulation of both metric and tetrad gravity is
well defined in globally hyperbolic space-times which are
asymptotically flat with suitable boundary conditions at spatial
infinity so that the only existing asymptotic symmetries are
associated with the ADM Poincare' generators. There are no Killing
vectors and the Dirac Hamiltonian turns out to be the weak ADM
energy.

\subsubsection{The Foldy-Wouthuysen Transformation for Spinning
Particles in an External Tetrad Field.}

In Ref.\cite{37} there was no study of a pseudo-classical FW
transformation along the lines of Ref. \cite{27}. Therefore,
following Ref.\cite{12}, we outline some original results of a novel study
involving a spinning particle
coupled to an external  cotetrad field \cite{38}. We start  from the
Lagrangian of Ref.\cite{37} coupled to an external tetrad field
[$(A)$ are flat indices]

\bea
 L(\tau ) &=& - {i\over 2}\, \xi_5\, {\dot \xi}_5 - {i\over 2}\,
 \eta_{(A)(B)}\, \xi^{(A)}\, \Big( {\dot \xi}^{(B)} + \omega_{\mu\,
 (C)}^{(B)}\, {\dot x}^{\mu}\, \xi^{(C)}\Big) -\nonumber \\
 &&{}\nonumber \\
 &-& mc\, \sqrt{\eta_{(A)(B)}\, \Big(E^{(A)}_{\mu}(x)\, {\dot x}^{\mu}
 - {i\over {mc}}\, \xi^{(A)}\, {\dot \xi}_5\Big)\, \Big(E^{(B)}_{\nu}(x)\,
 {\dot x}^{\nu} - {i\over {mc}}\, \xi^{(B)}\, {\dot
 \xi}_5\Big)}.\nonumber \\
 &&{}
 \label{4.1}
 \eea

Here $E^{(A)}_{\mu}(x)$ and $E^{\mu}_{(A)}(x)$ are cotetrad and
tetrad fields, respectively, evaluated at the particle position  and
satisfying
$$ E^{\mu}_{(A)}\,
E^{(A)}_{\nu} = \delta^{\mu}_{\nu}, \quad E^{(A)}_{\mu}\,
E^{\mu}_{(B)} = \delta^{(A)}_{(B)}, \quad E^{\mu}_{(A)}\,
{}^4g_{\mu\nu}\, E^{\nu}_{(B)} = \eta_{(a)(B)}.$$
 They describe an
external gravitational field, whose 4-metric is ${}^4g_{\mu\nu}(x) =
E^{(A)}_{\mu}(x)\, \eta_{(A)(B)}\, E^{(B)}_{\nu}(x)$, as a theory of
dynamical time-like observers endowed with spatial triads. The flat
limit corresponds to $E^{(A)}_{\mu} \rightarrow \delta^{(A)}_{\mu}$.

The spin connection is

\beq
 \omega_{\mu\, (B)}^{(A)} = E^{(A)}_{\alpha}\, \nabla_{\mu}\,
 E^{\alpha}_{(B)} = E^{(A)}_{\alpha}\, \Big(\partial_{\mu}\, E^{\alpha}_{(B)}
 + {}^4\Gamma^{\alpha}_{\mu\nu}\, E^{\nu}_{(B)}\Big).
 \label{4.2}
 \eeq

In phase space, after the elimination of the second class
constraints, we remain with the variables $x^{\mu}$, $p_{\mu}$,
$\xi^{(A)}$, $\xi_5$ satisfying the Dirac brackets $ \{ x^{\mu},
p_{\nu}\}^* =
 - \delta^{\mu}_{\nu}$, $\{ \xi^{(A)}, \xi^{(B)} \}^* = i\, \eta^{(A)(B)}$,
 $\{ \xi_5, \xi_5 \}^* = - i$. There are the following two first
 class constraints [$\omega_{(A)(B)(C)}$ are the Ricci rotation coefficients]

 \bea
   \chi &=& {}^4g^{\mu\nu}\, P_{\mu}\, P_{\nu} - m^2c^2 =
 \eta^{(A)(B)}\, P_{(A)}\, P_{(B)} - m^2c^2 \approx 0,\nonumber \\
 \chi_D &=& P_{\mu}\, E^{\mu}_{(A)}\, \xi^{(A)} - mc\, \xi_5 =
 P_{(A)}\, \xi^{(A)} - mc\, \xi_5 \approx 0,\nonumber \\
 &&{}\nonumber \\
 &&P_{\mu} = p_{\mu} - {i\over 2}\, \omega_{\mu(A)(B)}(x)\,
 \xi^{(A)}\, \xi^{(B)},\nonumber \\
 &&P_{(A)} = E^{\mu}_{(A)}(x)\, P_{\mu} = p_{(A)} -
 {i\over 2}\, \omega_{(A)(B)(C)}(x)\, \xi^{(B)}\, \xi^{(C)}.
 \label{4.3}
 \eea

They satisfy the algebra $\{ \chi_D, \chi_D \}^* = i \chi$, $ \{
\chi , \chi_D \}^* = \{ \chi , \chi \}^* = 0$.

The Dirac Hamiltonian is $H_D = \lambda (\tau )\, \chi +
\lambda_{\xi}(\tau )\, \chi_D$, where $\lambda (\tau )$ and
$\lambda_{\xi}(\tau )$ are Dirac multipliers, even and odd
respectively. As shown in Ref.\cite{27}, in Minkowski space-time in
the gauge $x^o \approx \tau$ the Dirac Hamiltonian becomes $H_D =
p_o + \tilde \lambda\, \xi^{(o)}\, \chi_D$ with a constant $\tilde
\lambda$: if we choose $\tilde \lambda = 2/\hbar$ the quantization
of the Grassmann variables produces the Hamiltonian $H = \vec \alpha
\cdot \vec p + \beta\, m$ associated with the free Dirac equation.
The FW transformation transforms it to $H^{'} = \beta\, \sqrt{m^2c^2
+ {\vec p}^2}$.

Since in the free case we have $\chi_D \mapsto \gamma_5\, (p_{\mu}\,
\gamma^{\mu} - mc)$, the FW transformation sends $\chi_D = p_{\mu}\,
\xi^{\mu} - mc\, \xi_5$ into $\chi_D^{'} = p_o\, \xi^o - mc\,
\xi_5$, which gives rise to the quantum Hamiltonian $\gamma_5\,
H^{'} = \gamma_5\, \beta\, \sqrt{m^2c^2 + {\vec p}^2}$.

These results are based on the fact that after quantization in the
Dirac-Pauli representation for the Dirac matrices $\xi^{(o)}$ and
$\xi_5$ become anti-diagonal matrices, while $\xi^{(a)}$ and
$\xi^{(o)}\, \xi_5$ become diagonal.

Therefore, the Dirac constraint can be written in the form

\bea
 \chi_D &=& P_{(A)}\, \xi^{(A)} - mc\, \xi_5 = p_{(o)}\, \xi^{(o)}
 - mc\, \xi_5 - {\cal E}_5(\xi ) - O_5(\xi ),\nonumber \\
 &&{}\nonumber \\
 {\cal E}_5(\xi ) &=& - {i\over 2}\, \omega_{(o)(b)(c)}\,
 \xi^{(b)}\, \xi^{(c)}\, \xi^{(o)} - i\, \omega_{(a)(o)(b)}\,
 \xi^{(o)}\, \xi^{(b)}\, \xi^{(a)},\nonumber \\
 &&{}\nonumber \\
 {\cal O}_5(\xi ) &=& - {\tilde P}_{(a)}\, \xi^{(a)} = - [p_{(a)}
 - {i\over 2}\, \omega_{(a)(b)(c)}\, \xi^{(b)}\, \xi^{(c)}]\,
 \xi^{(a)},
 \label{4.4}
 \eea

\noindent with the even ${\cal E}_5(\xi )$ and odd ${\cal O}_5(\xi
)$ parts going to diagonal and anti-diagonal matrices respectively
in the final quantum Hamiltonian $H^{'}$.

We look for a FW transformation such that the new odd part ${\cal
O}^{'}_5(\xi )$ is of the order $O({1\over {mc}})$. Then one could
devise an iterative scheme of FW transformations such that at the
$n-th$ step the odd part becomes of order $O({1\over {(mc)^n}})$.

Preliminary calculations \cite{38} for the pseudo-classical FW
canonical transformation $\chi_D \mapsto \chi^{'}_D = e^{\{ .,
s\}}\, \chi_D$ have been done. It turns out that the generating
function $S$ has the form [$S^{(a)} = {i\over 2}\,
\epsilon^{(a)}{}_{(b)(c)}\, \xi^{(b)}\, \xi^{(c)}$ is the spin
vector; we work in the Schwinger time-gauge with tetrads adapted to
the \lq\lq 3+1" splitting of space-time]

\bea
  S &=& - 2i\, P_{(a)}\, \xi^{(a)}\, \xi_5\, \theta (\alpha ),
  \nonumber \\
 &&{}\nonumber \\
  \theta (\alpha ) &=& {1\over {2\, \sqrt{\alpha}}}\, arctg\,
  {{\sqrt{\alpha}}\over {mc}}, \nonumber \\
 &&{}\nonumber \\
  \alpha &=& i\, \{ P_{(a)}\, \xi^{(a)}, P_{(b)}\, \xi^{(b)} \}^* =
 \nonumber \\
 &=& - \eta^{(a)(b)}\, p_{(a)}\, p_{(b)} + p_{(a)}\, \eta^{(a)(b)}\,
  \omega_{(b)(c)(d)}\, \epsilon^{(c)(d)}{}_{(e)}\,  S^{(e)}
  +\nonumber \\
  &+& 2 p_{(o)}\, \omega_{(a)(b)(o)}\,  \epsilon^{(a)(b)}{}_{(c)}\, S^{(c)}
 + 2i\, \xi^{(o)}\, \xi^{(a)}\, \Big[\omega_{(o)(a)(o)}\,
  \omega_{(b)(c)(o)}\, \epsilon^{(b)(c)}{}_{(d)}\, S^{(d)}
  +\nonumber \\
  &+& \Big(p_{(b)} - {1\over 2}\,
  \omega_{(b)(c)(d)}\, \epsilon^{(c)(d)}{}_{(e)}\, S^{(e)}
  \Big)\, \eta^{(b)(c)}\, \Big(\omega_{(a)(c)(o)} -
  \omega_{(c)(a)(o)}\Big) \Big].
  \label{4.5}
  \eea

The final form of $\chi_D^{'}$ is

\bea
 &&p_{(o)}\, \xi^{(o)} - {\cal E}_m\, \xi_5 - {\cal E}_{(s)5}(\xi )
 - {\cal O}_{(s)5}(\xi ) \approx 0,\nonumber \\
 &&{}\nonumber \\
 &&{}\nonumber \\
  {\cal E}_m &=& \sqrt{m^2c^2 - \eta^{(a)(b)}\, p_{(a)}\, p_{(b)}} +
 {{p_{(a)}\, \eta^{(a)(b)}\, \omega_{(b)(c)(d)}\, \epsilon^{(c)(d)}{}_{(e)}\,
 S^{(e)}}\over {2\, \sqrt{m^2c^2 - \eta^{(a)(b)}\, p_{(a)}\, p_{(b)}}}},
 \nonumber \\
 &&{}\nonumber \\
 {\cal E}_{(s)5}(\xi ) &=& {1\over 2}\, \omega_{(o)(b)(c)}\, \epsilon^{(b)(c)}{}_{(d)}\, S^{(d)}
 \xi^{(o)} + {{p_{(o)}\, {\buildrel o \over
\omega}_{(a)(o)(b)}\, \epsilon^{(a)(b)}{}_{(c)}\, S^{(c)}}\over
{\sqrt{m^2c^2 + \alpha_p}}}\, \xi_5 +\nonumber \\
 &+& {{mc - \sqrt{m^2c^2 + \alpha_p}}\over {\alpha_p\,
\sqrt{m^2c^2 + \alpha_p}}}\, p_{(m)}\, \Big(p_{(o)}\,
 {\buildrel o \over
\omega}_{(o)(o)(a)} +\nonumber \\
 &&{}\nonumber \\
&&+ p_{(b)}\, \eta^{(b)(c)}\,
 {\buildrel o \over \omega}_{(a)(o)(c)}\Big)\,
 \epsilon^{(a)(m)}{}_{(n)}\, S^{(n)}\, \xi^{(o)},\nonumber \\
 &&{}\nonumber \\
 {\cal O}_{(s)5}(\xi ) &=& O({1\over {mc}}).
 \label{4.6}
 \eea

The mass shell constraint is $\chi^{'} = - i\, \{ \chi_D^{'},
\chi_D^{'} \}$ and the new Dirac Hamiltonian is $H^{'}_D = {{d
S}\over {d\tau}} + \lambda (\tau )\, \chi^{'} + \lambda_{\xi}(\tau
)\, \chi_D^{'}$.

Once these calculations will be completed  and the quantization of the
pseudo-classical FW transformation will be done, we will have a
first control on the coupling of positive-energy $({1\over 2}, 0)$
2-spinors to an arbitrary external gravitational field in the
framework of tetrad gravity. Since due to Ref.\cite{10,12} it is
clear how to disentangle the inertial effects (gauge variables) from
the genuine tidal effects (the two pairs of canonically conjugate
Dirac observables, becoming the spin-2 degrees of freedom of the
weak field approximation), it will be possible to study the problem
of the spin-rotation couplings in gravity \cite{1,2,3,4} in the
general case of no Killing vectors and the reliability of the use of
the equivalence principle in this area.

Till now this topic has been treated only in Schwartzschild, Kerr
and Post-Newtonian spherically symmetric space-times (they are
enough for applications inside the solar system) but only in the 1+3
point of view. The best results for the quantum FW transformation in
this special class of space-times have been obtained in
Refs.\cite{39}, after the block-diagonalization of the Dirac
Hamiltonian with the Eriksen-Korlsrud method (inequivalent to the FW
transformation already in the free case) done in Ref.\cite{40},
which had shown the presence of a dipole spin-gravity coupling
absent in Minkowski accelerated frames \cite{36}.

The main results of Ref.\cite{39} in a weak spherically symmetric
gravitational field is the absence of the precession of spin of
fermions at rest (the spin rotation is the de Sitter one like for a
classical gyroscope) and the validity of the equivalence principle
understood as minimal coupling of fermions to gravity. However
different observables have different behaviors: i) the helicity of a
massive Dirac particle has the same evolution in these gravitational
fields and in accelerated frames (in the 1+3 point of view); ii) the
spin and the momentum (defining the helicity motion) rotate in the
same direction but with different angular velocities (in the
semiclassical limit $\omega_{spin} - \omega_{momentum} = {m\over
{{\vec p}^2}}\, \vec g \times \vec p$ with $\vec g = - {{G\, M}\over
{r^3}}\, \vec r$), which differ from the ones in accelerated frames
by kinematical corrections (the spin precession is 3 times bigger).

In the second paper of Ref.\cite{39}, it is said that the Newtonian
equivalence principle (equality of inertial and gravitational
masses) and its Post-Newtonian version (absence of gravitational
analogs of electric dipole and anomalous magnetic moments) are till
now well tested. Then there is the following version of the
Post-Newtonian equivalence principle: the equality of the
frequencies of precession of quantum (spin) and classical (orbital)
angular momenta and the preservation of the helicity of a Dirac
particle in a gravito-magnetic field. In this paper \cite{39} and in
Ref. \cite{3} there are proposals of future experiments to test
which point of view is the more suitable to describe this important
area of gravitational effects.

\subsubsection{The Dirac Equation in Tetrad Gravity}

Finally the study started \cite{41} of tetrad gravity coupled to the
Dirac field to find the full set of constraints both for gravity and
for the Dirac field as matter to extend the results of Ref.\cite{35}
to curved space-times.

When this work will be completed, one will be able to study the FW
transformation in general space-times along the lines of
Ref.\cite{39} for the special class of space-times with Killing
symmetries without relying on the pseudo-classical approximation.
This will give the most general answer to the problem of
spin-gravity couplings both in the solar system and in astrophysical
contexts. In particular one will be able to see whether the results
of the 1+3 point of view about massless fermions (neutrinos)
\cite{42} will be confirmed.

\section{Conclusions}

In this paper we  have reviewed and discussed the most important
aspects of the problem of the spin-rotation couplings, passing from
special to general relativity as well as from classical to
semiclassical and quantum physics. The present status of the
research and an original presentation of material otherwise
scattered into a large number of papers is discussed. All the
theoretical expectations and difficulties concerning the existence
of a spin-rotation coupling, as foreseen by Mashhoon long ago, are
framed both into the \lq\lq 1+3" point of view and into a more
general context involving \lq\lq 3+1" space-time splitting
techniques.

In this way we have been looking at the status of the art as well as
at the open issues and the way to proceed. Due to the lack of
experimental data, it is not yet clear which of the two points of
view is more suitable to describe physics on the Earth and inside
the solar system. At the classical level many of the complications
introduced by the 3+1 view may be negligible, but at the quantum
level the presence of time-dependent unitary transformations may
change the situation modifying the coupling present in some 1+3
approaches.

In conclusion, it would important to understand better the
transition from the \lq\lq 1+3" view to the \lq\lq 3+1" one. A first
tool has been introduced in Ref.\cite{9}, where it is shown that any
non-surface-forming congruence of time-like observers with unit
4-velocity field $u^{\mu}$ of the 1+3 point of view can be
reinterpreted as a congruence in which $u^{\mu}(z(\tau ,\sigma^r)) =
\partial\, z^{\mu}(\tau ,\sigma^r)/\partial\, \tau /
\sqrt{g_{\tau\tau (\tau ,\sigma^r)}}$ for some M$\o$ller admissible
embedding $z^{\mu}(\tau ,\sigma^r)$ defining a good clock
synchronization convention on instantaneous 3-spaces (
$\epsilon^{\mu}_A$ is an arbitrary orthonormal tetrad and $C(\tau )$
is a function sufficiently small so that M$\o$ller conditions are
satisfied)

 \bea
  z^{\mu}(\tau ,\vec \sigma ) 
  &=& x^{\mu}(\tau ) +  \epsilon^{\mu}_r\, \sigma^r +
  \int_o^{\tau} d\tau_1\, C(\tau_1)\, \epsilon_{\tau \nu}\,
  \Big[u^{\nu}(\tau_1, \vec \sigma )\, u^{\mu}(\tau_1, \vec \sigma )\nonumber  \\
  && - u^{\nu}(\tau ,\vec 0)\, u^{\mu}(\tau ,\vec 0)\Big].
  \label{V10}
  \eea

As a consequence, given any congruence associated with nonzero vorticity, we can find admissible \lq\lq 3+1" splittings of Minkowski
space-time, with the space-like simultaneity  leaves not 
in general not orthogonal to the reference world line $x^\mu (\tau)$ chosen as the origin, which allow to define genuine
instantaneous 3-spaces with synchronized clocks.

\section*{Acknowledgments}

DB acknowledges ICRANet for support.

\end{document}